\documentclass[useAMS,usenatbib]{mn2e}

\usepackage[T1]{fontenc}
\usepackage{pslatex}
\usepackage{amsmath}
\usepackage{amsfonts}
\usepackage{color}
\usepackage{url}
\usepackage{graphicx}
\usepackage{subfigure}
\usepackage{amssymb}
\usepackage{soul}
\usepackage{booktabs}
\usepackage{array}
{\newif\ifnotend
\notendtrue
\def\veclist{ABCDEFGHIJKLMNOPQRSTUVWXYZabcdefghijklmnopqrstuvwxyz.}
\def\top#1#2.{#1}
\def\tail#1#2.{#2.}
\loop\expandafter\xdef\csname v\expandafter\top\veclist\endcsname%
{{\noexpand\bf\expandafter\top\veclist}}
\edef\veclist{\expandafter\tail\veclist}
\if\veclist.\notendfalse\fi\ifnotend\repeat}
\def\pa{\partial}

\mathchardef\mhyphen="2D

\title[Gas flow in barred potentials III]{Gas flow in barred potentials - III. Effects of varying the Quadrupole.}
\author[Sormani, Binney \& Magorrian]{Mattia C. Sormani, James Binney and John Magorrian\\
$^1$ Rudolf Peierls Centre for Theoretical Physics, 1 Keble Road, Oxford
OX1 3NP}

\begin{document}

\date{}

\def\p{\partial}
\def\Omegap{\Omega_{\rm p}}

\newcommand{\di}{\mathrm{d}}
\newcommand{\bfx}{\mathbf{x}}
\newcommand{\bfe}{\mathbf{e}}
\newcommand{\vlos}{\mathrm{v}_{\rm los}}
\newcommand{\Tspin}{T_{\rm s}}
\newcommand{\Tb}{T_{\rm b}}
\newcommand{\degree}{\ensuremath{^\circ}}
\newcommand{\Th}{T_{\rm h}}
\newcommand{\Tc}{T_{\rm c}}
\newcommand{\bfr}{\mathbf{r}}
\newcommand{\bfv}{\mathbf{v}}
\newcommand{\pc}{\,{\rm pc}}
\newcommand{\kpc}{\,{\rm kpc}}
\newcommand{\Myr}{\,{\rm Myr}}
\newcommand{\Gyr}{\,{\rm Gyr}}
\newcommand{\kms}{\,{\rm km\, s^{-1}}}
\newcommand{\de}[2]{\frac{\partial #1}{\partial {#2}}}
\newcommand{\cs}{c_{\rm s}}
\newcommand{\rb}{r_{\rm b}}
\newcommand{\rqu}{r_{\rm q}}

\maketitle

\begin{abstract}
 We run hydrodynamical simulations of a 2D isothermal non
 self-gravitating inviscid gas flowing in a rigidly rotating
 externally imposed potential formed by only two components: a
 monopole and a quadrupole.
 We explore systematically the effects of varying the quadrupole
 while keeping fixed the monopole and discuss the consequences for
 the interpretation of longitude-velocity diagrams in the Milky Way.
 We find that the gas flow can constrain the quadrupole of the
 potential and the characteristics of the bar that generates it.
 The exponential scale length of the bar must be at least $1.5\kpc$.
 The strength of the bar is also constrained. Our global
 interpretation favours a pattern speed of $\Omega=40\, \kms
 \kpc^{-1}$.
 We find that for most observational features, there exist a value of
 the parameters that matches each individual feature well, but is
 difficult to reproduce all the important features at once.
 Due to the intractably high number of parameters involved in the
 general problem, quantitative fitting methods that can run automatic
 searches in parameter space are necessary.
\end{abstract}


\section{Introduction} \label{sec:introduction}
It is now well-established that the Milky Way is a barred galaxy (e.g.
\citealt[]{Stanek1994,Dwek1995,Binney1997,WeggGerhard2013}, see also
\citealt{FuxBarReview,GerhardWeggReview2014} for reviews). Many authors have
modelled the gas flow in the Milky way and compared it with the available
data \citep{MulderLiem1986,Binney++1991,JenkinsBinney94,weinersellwood1999,
EnglmaierGerhard1999,Lee++1999,Fux1999,Bissantz++2003,RFC2008,Baba++2010,
Pettitt++2014,Pettitt++2015}.  However, a number of issues remain unresolved
and the data still contain a wealth of information capable of constraining
the Galactic potential that has yet to be fully
exploited by gas dynamical models \citep{SM15}.

In two previous papers \citep[][hereafter SBM15a and
SBM15b]{SBM2015a,SBM2015b} we investigated the dynamics of gas flow in barred
potentials. We studied in detail the relation between the flow of a 2D
isothermal non self-gravitating inviscid gas and closed ballistic orbits in
the same externally imposed rigidly rotating barred potential. We found that
hydro streamlines closely follow ballistic closed orbits at large and small
radii, and tiny deviations of the hydro streamlines from ballistic closed
orbits generate bar-driven spiral arms as kinematic density waves. At
intermediate radii shocks arise and the streamlines shift between two
families of closed orbits. We showed that the sound speed in the gas and the
spatial resolution of the grid both affect the gas flow  significantly. 

In SBM15a we discussed the implications of our simulations for the
interpretation of longitude-velocity diagrams (hereafter $(l,v)$ diagrams) of
atomic (HI) and molecular (CO,CS) gas in the Milky Way. The simulations
presented in SBM15a were based on the \cite{Binney++1991} barred potential,
which was originally used to construct a picture of the flow of gas through
the central few kiloparsecs of our Galaxy on the assumption that gas follows
closed orbits. Our simulations provided strong support for this assumption,
but refined the \cite{Binney++1991} picture of gas flow in several respects.
Specifically: (i) In \cite{Binney++1991} the parallelogram in the $(l,v)$ plot
for CO was interpreted as the trace of the cusped orbit, while we found that
the shocks form two sides of the CO parallelogram, and conjectured that the
prominence of the CO parallelogram is due to efficient conversion of atomic
gas into molecular gas. (ii) \cite{Binney++1991} did not have an explanation
for the observed asymmetry in the distribution of molecular emission near the
Galactic centre, while we argued that a promising explanation for the
asymmetry is provided by the way the \emph{wiggle instability}
\citep{WadaKoda2004,KimKimKim2014} makes the flow through the shocks
unstable. The large fluctuations generated by this instability might 
cause the conversion efficiency to fluctuate wildly and give rise to gross
asymmetry in the distribution of molecular gas.

However, since we did not keep track of the chemistry of the ISM,
items (i) and (ii) remained merely promising conjectures. 

SBM15a identified two key features of the observed $(l,v)$ diagrams that were
still unexplained after their revision of the \cite{Binney++1991} model: (i)
Coherent broad features like the $3\kpc$ arm and its counterpart on the far
side of the Galaxy \citep{Dame2008} -- these where absent from the
simulations of SBM15a. (ii) Forbidden emission at large longitudes -- in the
SBM15a models forbidden emission covers a significantly smaller portion of
the $(l, v)$ diagram than in the data.

SBM15a suggested two main directions for improving their models: First,
inclusion of a law for the conversion of gas between atomic and molecular
forms, so when gas is compressed at a shock much of it is converted to
molecular gas. Second, modification of
the quadrupole moment of the bar, since a higher quadrupole moment should
generate stronger spiral arms and stronger non-circular motions, which are
the likely explanations for internal features and forbidden velocity emission
in the observational $(l,v)$ diagrams. SBM15b explored the mechanism by which
the bar generates the spirals that are responsible for internal structure in
the $(l,v)$ plane. Consequently, we here implement the second upgrade
recommended by SBM15a by systematically exploring the
effects on the gas flow of a variation of the quadrupole component of the
potential, and discuss the consequence of this variation for the
interpretation of Milky Way $(l,v)$ diagrams. 

In Section~\ref{sec:methods} we explain our numerical methods. In
Section~\ref{sec:observations} we enumerate the bar's signatures in the
$(l,v)$ plane. In Section~\ref{sec:archetypal} we describe a reference model,
which we use to relate structures in the galactic plane to those in the
$(l,v)$ plot. In Section~\ref{sec:quadrupole} we study how the structure of
the $(l,v)$ plane changes as we change the parameters defining the bar. In
Section~\ref{sec:implications} we outline the extent to which the Galactic
bar is constrained by the observed $(l,v)$ diagrams, and in
Section~\ref{sec:conclusion} we sum up and consider promising directions for
future work.

\section{Methods} \label{sec:methods}
\subsection{The potential} \label{sec:potential}
We assume that the gas flows in a simple externally imposed
two-dimensional barred potential that rotates at constant pattern speed
$\Omega_{\rm p}$. The potential can be expanded in multipoles
\begin{equation}
\Phi(R,\phi) = \Phi_0(R) +  \sum_{m=1}^\infty \Phi_m(R) \cos(m \phi +
\phi_m) \, ,
\end{equation}
where $\{R,\phi\}$ are planar polar coordinates, $\phi_m$ are constants and
$\Phi_m$ are functions of $R$ only. We assume that the potential comprises
only monopole and quadrupole terms, so
\begin{equation}
\Phi(R,\phi) = \Phi_0(R) + \Phi_2(R) \cos(2 \phi) \, .
\end{equation}

\begin{figure}
\includegraphics[width=0.5\textwidth]{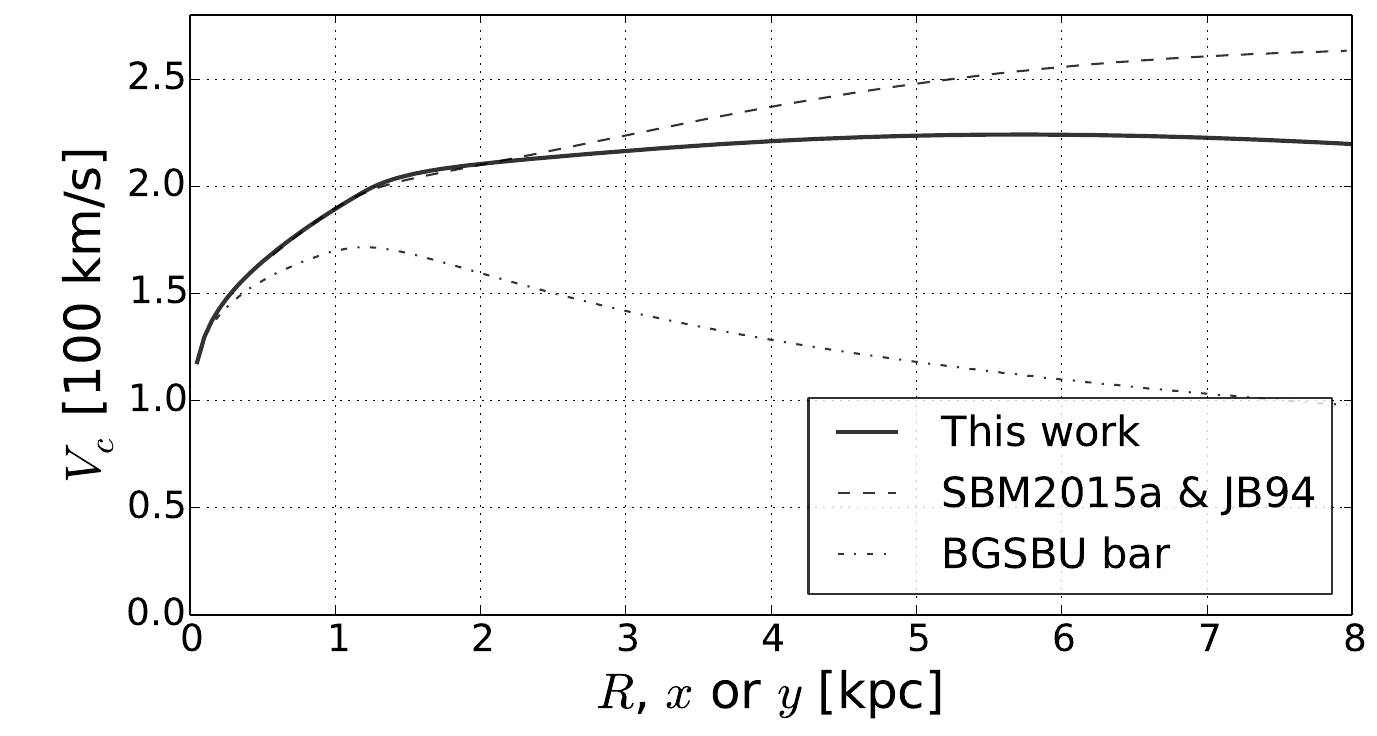}
\caption{Circular speeds. Solid Line: potential used in this paper. Dashed line: potential used in \protect\cite{JenkinsBinney94,SBM2015a}. Dot-dashed line: potential used in \protect\cite{Binney++1991}. All the curves in this picture are calculated from the monopole component of each potential.} 
\label{fig:Phi0}
\end{figure}

The solid line in Fig.~\ref{fig:Phi0} shows the circular speed $v_c$ implied
by the monopole $\Phi_0(r)$ we adopt in the present paper, while the dashed line shows $v_c$ for the monopole
used by \cite{JenkinsBinney94} and SBM15a. The latter monopole is unrealistic
at large $R$ because its circular speed becomes excessive, whereas the
monopole used here coincides with the monopole of \cite{SBM2015a} at small
radii but at greater radii generates a circular speed that plateaus at
$220\kms$.  The dot-dashed line shows for comparison the circular speed of
the \cite{Binney++1991} potential. The potential used in
\cite{JenkinsBinney94} and SBM2015a differs from the potential used in
\cite{Binney++1991} only by the addition of an axisymmetric disc component.

Our quadrupole $\Phi_2$  is generated by the density
distribution 
\begin{equation}
\rho_2(r,\phi,\theta) = \frac{K A}{\rqu^2}  
\exp \left( - \frac{2 r}{\rqu} \right) \sin^2\theta\, \cos(2 \phi) \;, \label{eq:rho2}
\end{equation}
where $\{r,\theta,\phi \}$ are spherical coordinates with the $\theta=0$ axis
pointing towards the north galactic pole. The constants are\footnote{\emph{Note added after publication:} this factor has been corrected. The published version contains a typo.}
\begin{equation}
K = \frac{v_{0}^2}{4 \pi G} \exp\left(2\right) \;,
\end{equation}
with $v_0 = 220 \, \kms$ and $G$ is the gravitational constant.
The two main free parameters of
$\rho_2$ are the quadrupole strength $A$ and the quadrupole length $\rqu$. We
have chosen the form of the quadrupole density distribution to be exponential
as recent infrared photometry has found that the Milky Way bar density
profile is roughly exponential \citep[][]{WeggGerhard2013}.

Since $\rho_2$ is proportional to the real part of $Y_2^2$, which is an
eigenfunction of the Laplacian operator, the density distribution
\eqref{eq:rho2} gives rise to the potential
\begin{equation}
\Phi_2(r,\phi,\theta) = \Phi_2(R) \sin^2\theta\, \cos(2 \phi) \;.
\end{equation}
Our simulations are two-dimensional, so we only evaluate $\Phi_2$ in the
plane $\theta=\pi/2$. Thus, while the 3D density distribution given by
Eq.~\eqref{eq:rho2} should not be considered a realistic density distribution
for our Galaxy, the resulting 2D potential in the plane can be obtained by a
more realistic 3D density distribution. To be physical, a potential must come
from a non-negative density distribution. We have checked that when our
monopole is realised through a spherical symmetric distribution, the total
density distribution is positive for all the values of the parameters used in
this paper. 

\begin{figure}
\includegraphics[width=0.5\textwidth]{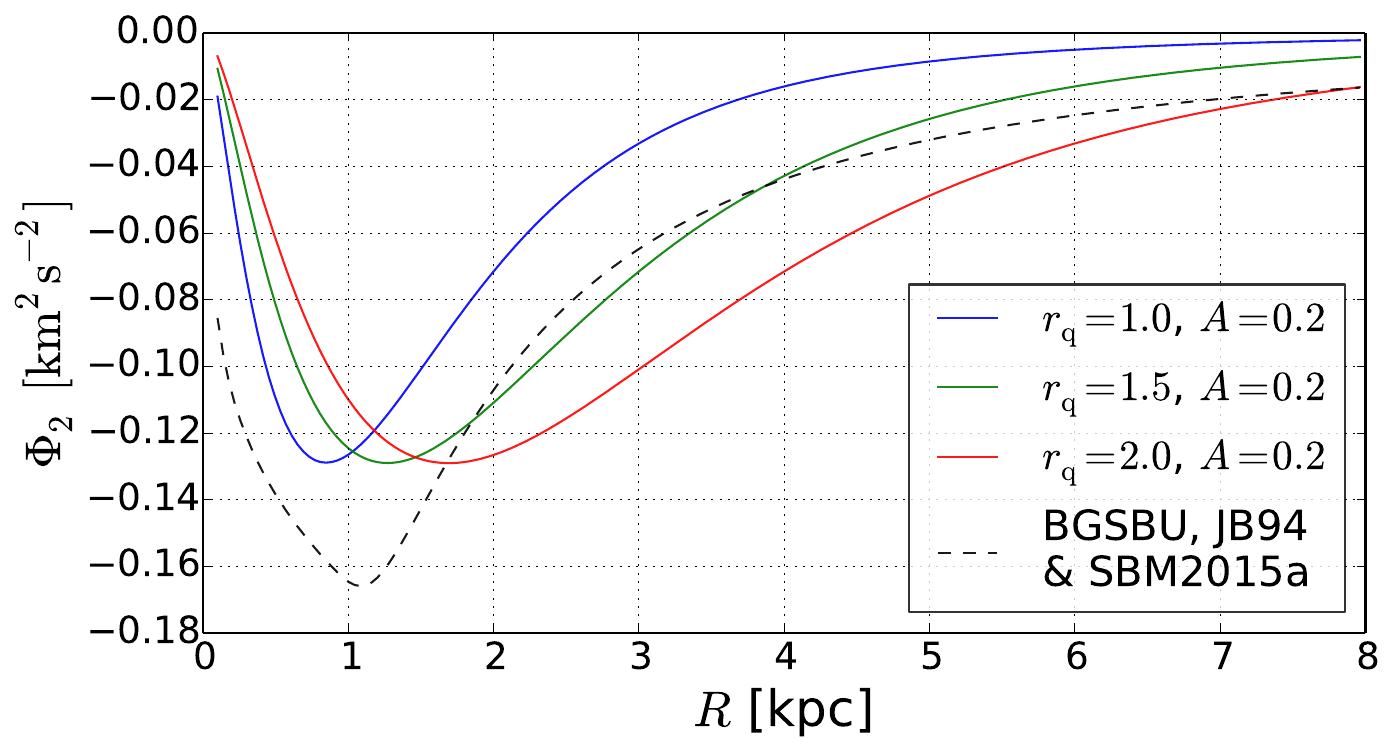}
\caption{The quadrupole component $\Phi_2$. In full lines, for the class of potentials used in this paper for the three different values of the bar length $\rqu$ and for a bar strength $A=0.2$. The curves for different values $A$ are not shown but can be easily obtained by scaling linearly the curves shown. In dashed line the quadrupole component of the bar used in \protect\cite{Binney++1991,JenkinsBinney94,SBM2015a} is shown for comparison. \emph{Note added after publication}: the $y$ label in this plot should be in units of $[100\rm km/s]^2$, not $[\rm km/s]^2$.} 
\label{fig:Phi2}
\end{figure}
\subsection{Hydro Simulation Scheme} \label{sec:hydro}
In our simulations, we assume that the gas is a two-dimensional inviscid
isothermal fluid governed by the Euler equations. An additional term is
introduced in the continuity equation to implement the recycling law of
\cite{Athan92b}.  The dynamical equations in an inertial frame are
\begin{equation} \begin{split}
	\pa_t \rho + \nabla \cdot \left( \rho \bfv \right) & = \alpha (\rho_0^2 - \rho^2),  \\
	\pa_t \bfv + \left( \bfv \cdot \nabla \right) \bfv & = - \frac{\nabla P}{\rho} -\nabla \Phi, \\
	P & = \cs^2 \rho,\,
\end{split} \end{equation}
where $\rho$ is the surface density of the gas, $P$ is the pressure, $\Phi$
is the gravitational potential, $\bfv$ is the velocity, $\cs$ is the sound
speed, $\alpha$ is a constant representing the efficiency of the recycling
law and $\rho_0$ is the initial surface density. In our simulations the gas
is assumed to be isothermal, hence the sound speed is a constant number which is assumed to be $\cs=10\kms$.

The recycling law was originally meant to take into account in a simple way
the effects of star formation and stellar mass loss. In practice, the only
effect of the recycling law is to prevent too much gas from accumulating in
the very centre and to replace gas lost at the boundary due to the outflow
boundary conditions. It does not affect the morphology of the results, so our
results do not change if we disable the recycling law. We adopt recycling
efficiency $\alpha=0.3 \, M_\odot \pc^{-2}\Gyr^{-1}$ and initial density
$\rho_0 = 1 \, M_\odot\pc^{-2}$.  

We use a grid-based, Eulerian code based on the second-order flux-splitting
scheme developed by \cite{vanAlbada+82} and later used by \cite{Athan92b},
\cite{weinersellwood1999} and others to study gas dynamics in bar potentials.
The implementation used here is the same as was used by SBM15a,b.

We used a grid $N \times N$ to simulate a square $20\kpc$ on a side, where
$N$ defines the resolution of the simulation. We start with gas in
equilibrium on circular orbits in an axisymmetrized potential and, to avoid
transients, turn on the non-axisymmetric part of the potential gradually during the first $615 \rm Myr$. 
We use outflow boundary conditions: gas can freely escape
the simulated region, after which it is lost forever. The potential well is
sufficiently deep, however, to prevent excessive quantities of material from
escaping. 

\subsection{Projecting to the $(l,v)$ plane} \label{sec:projection}
We adopt a very simple projection procedure to produce the predicted $(l,v)$
distributions for each simulation snapshot ($\rho(\bfx),\bfv(\bfx)$).
We assume that the Sun is undergoing circular motion
at a radius $R_0=8\kpc$ with speed $\Theta_0 = 220\kms$.  Calling $\phi$ the
angle between the major axis and the Sun--GC line, the Cartesian coordinates
of the Sun are given by $x_\odot = R_0 \cos \phi$, $y_\odot = R_0 \sin \phi$. 
In our models, we only project material inside the solar circle.

The resolution of our $(l,v)$ diagrams is $\Delta l=0.25\degree$ in longitude
and $\Delta v=2.5\kms$ in velocity.  Along each line of sight, we sample the
density and the velocity by linearly interpolating the results of the
simulations at points separated by $\delta s =1 \pc$.  These density measures
are accumulated in velocity bins of width $\Delta v=2.5\kms$.  The final
$(l,v)$ intensity at the chosen longitude in each range of velocity are obtained by summing over all the relevant 
points along the line of sight weighted by their masses.

This procedure yields a brightness temperature that is linear in
column density which is equivalent to the simplest radiative transfer
calculation \citep[e.g.,][Formula (8.17) and (8.20)]{BM}.  In the case of HI, the brightness temperature is linear in the
column density if the gas has constant spin temperature and its optical depth
is negligible.  So our projection is equivalent to simple HI radiative
transfer in the constant-temperature, optically-thin case.  The assumption of
constant temperature is known to be a simplification for Galactic HI, which
is instead often modelled as a medium made by two or more phases at different
temperatures \citep[see for example][]{Ferriere2001}.  In the case of
${}^{12}$CO, the brightness temperature is not linearly related to density
when considering a single cloud because molecular clouds are typically
optically thick at $2.6\,$mm. However,  a linear relationship will hold between
brightness temperature and the number density of unresolved CO clouds
provided the cloud density is low enough for shadowing of clouds to be
unimportant \citep[see, e.g.,][\S 8.1.4]{BM}.

\begin{figure}
\includegraphics[width=0.5\textwidth]{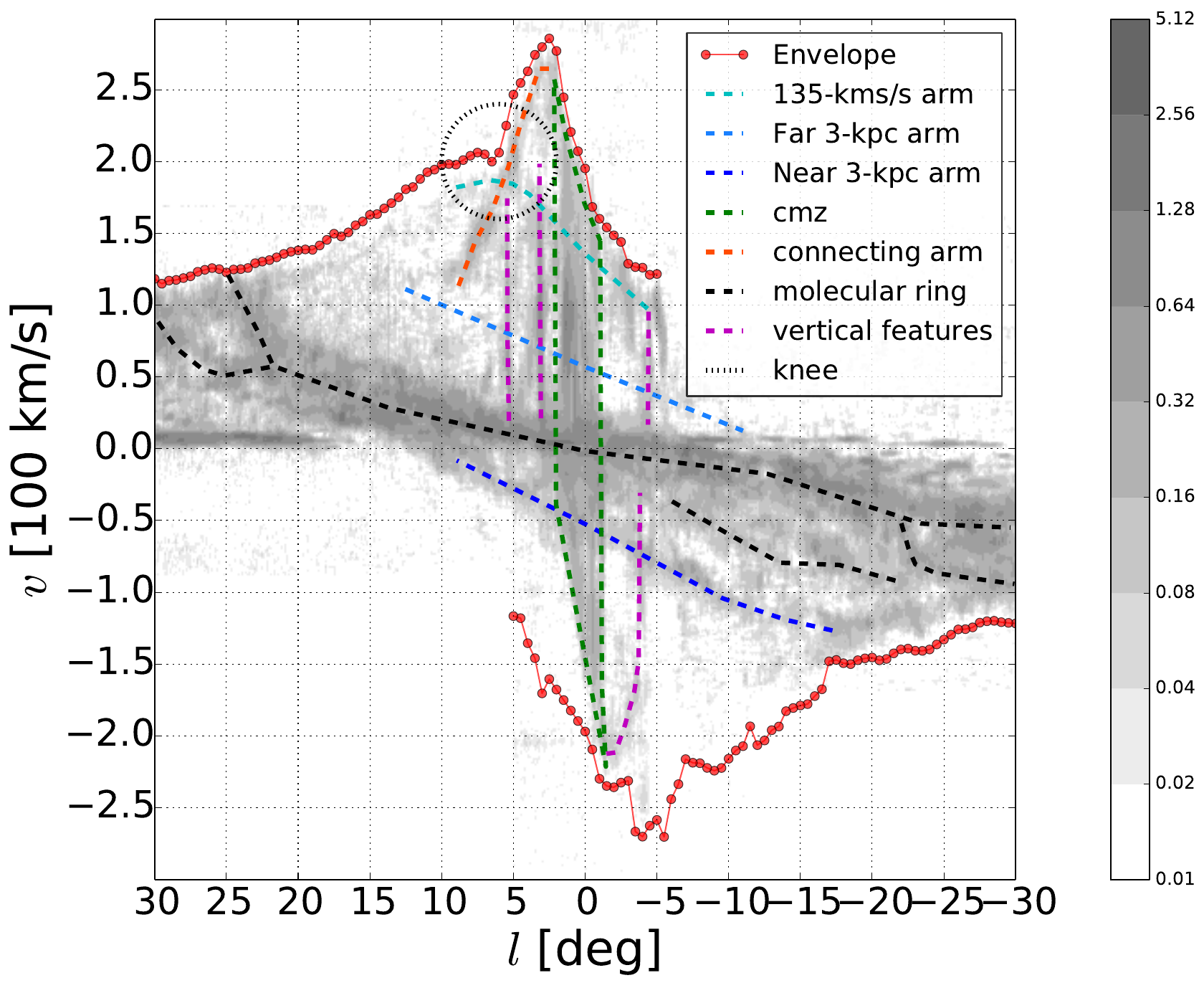}
\includegraphics[width=0.5\textwidth]{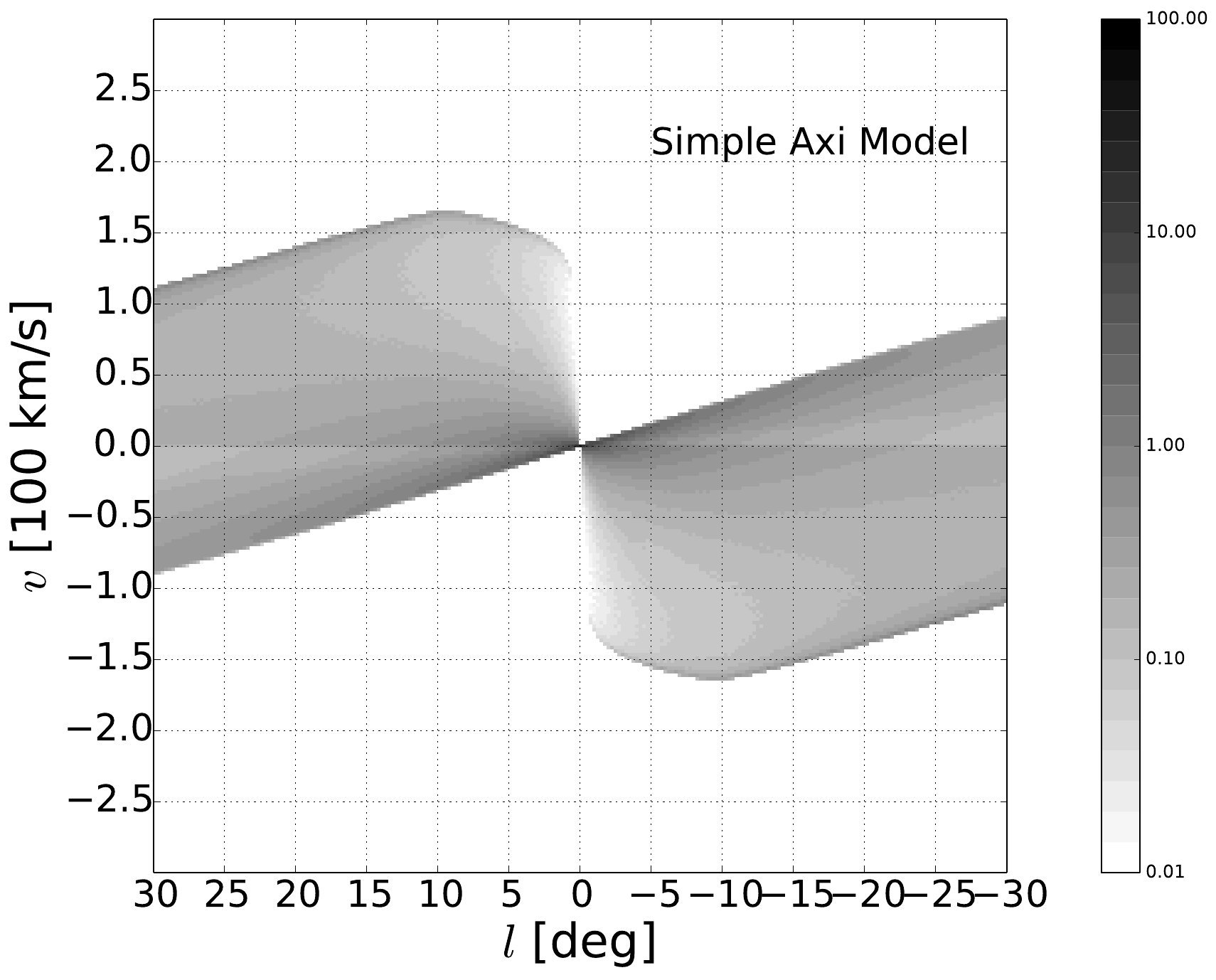}
\caption{Upper panel: CO observations with features superimposed. The red
dots trace the envelope determined from HI
data as explained in
\protect\cite{SM15}. Dashed lines trace internal features. The colorbar is in K. Lower panel:
appearance of the diagram in the absence of the quadrupole potential under
the assumption that the gas density is proportional to $\exp(-R / 2 R_0)$. The colorbar is in arbitrary units.}
\label{fig:obs2}
\end{figure}

\section{Signatures of the bar} \label{sec:observations}
\begin{figure}
\includegraphics[width=0.5\textwidth]{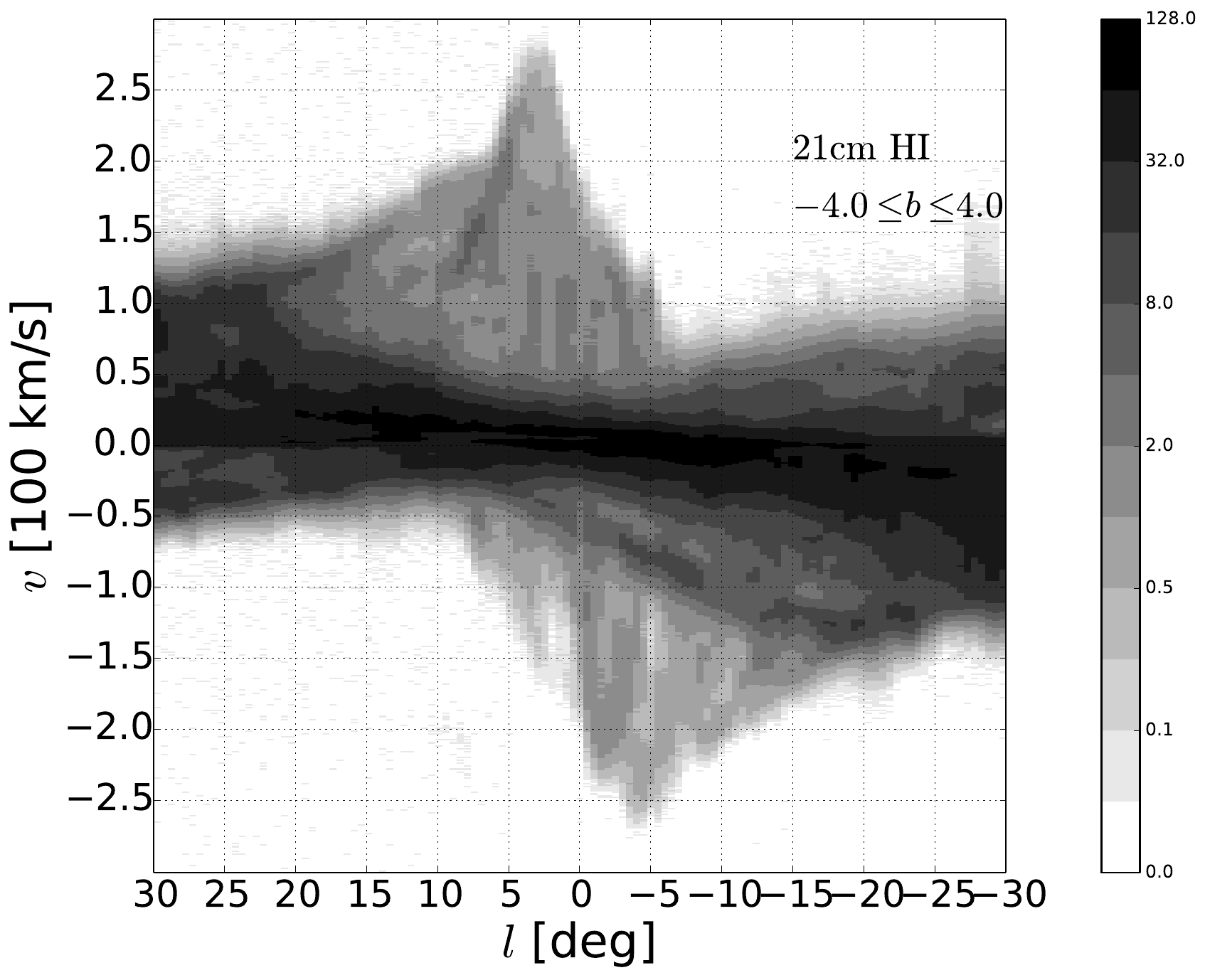}
\includegraphics[width=0.5\textwidth]{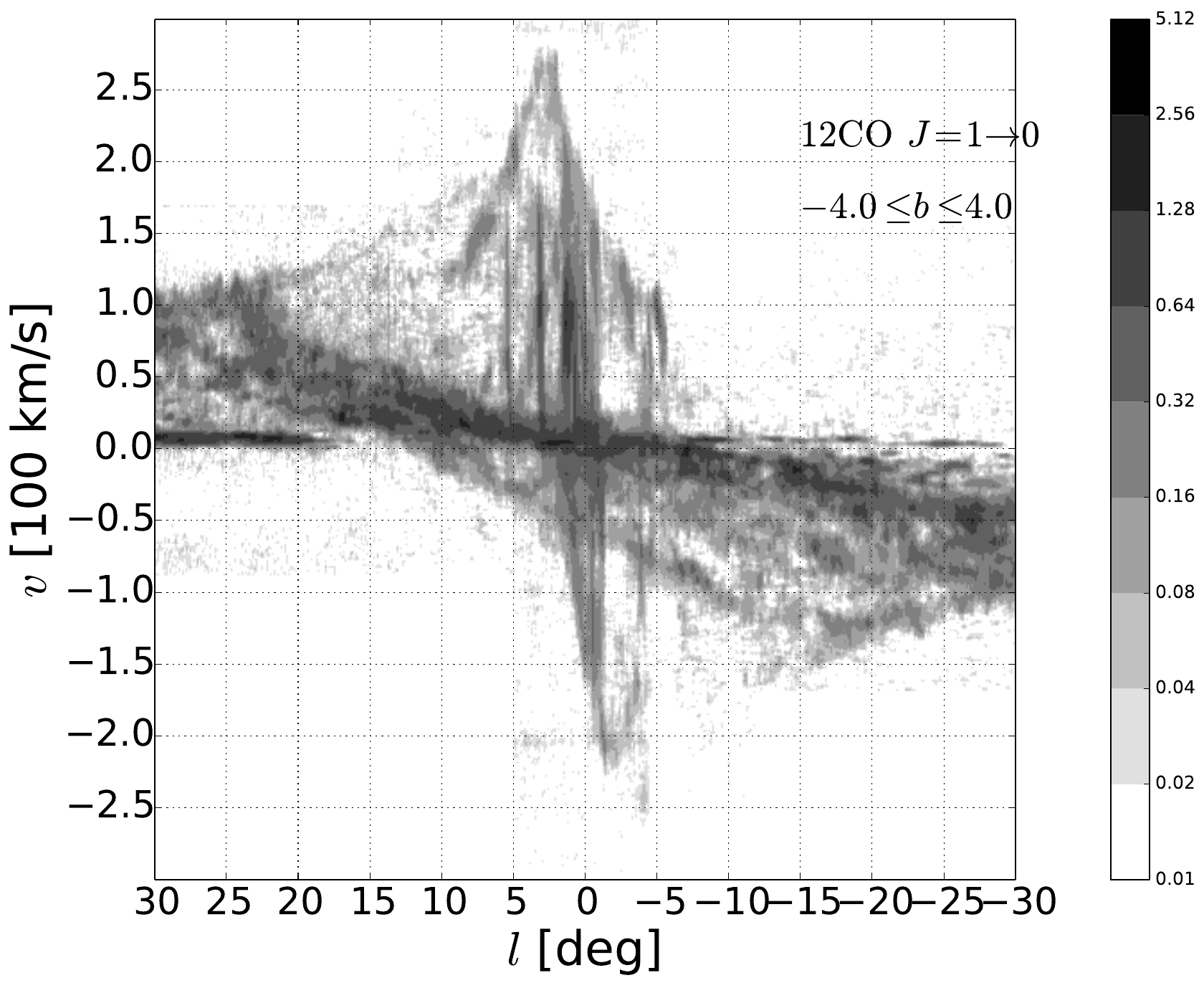}
\caption{Upper panel: HI observations integrated over $|b|<4\degree$. Lower panel: CO observations integrated over $|b|<4\degree$. Colorbars units are in K.} 
\label{fig:obs1}
\end{figure}

In the upper panel of Fig.~\ref{fig:obs2} the grey scale shows the intensity
in the $(l,v)$ plane of $2.6\,$mm ${\rm J=1\rightarrow 0}$ line radiation by CO at $|l|<30\degree$.
Coloured lines schematically trace features identifiable in the CO and 21-cm HI $(l,v)$ diagrams. 
The lower panel shows the very different distribution
of emission we would expect in the absence of the quadrupole
$\Phi_2$.  The latter has a big impact on $(l,v)$ diagrams and we explore how
the features marked in the upper panel can be used to constrain $\Phi_2$.  We
will in the process indicate where in the Galaxy the gas lies that gives rise
to most of these features.  The background to our discussion can be found in
\cite{burtonbook} and \cite{BM}.  Fig. \ref{fig:obs1} shows the
longitude-velocity diagrams for 21cm HI data \citep{HIdata} in the upper
panel and of $^{12}$CO data \citep{COdata} in the lower panel. 

The key
features of the observed $(l,v)$ plots are

\begin{enumerate} 

\item{\it Emission at forbidden velocities}. If gas everywhere moved on
circles, emission in the quadrants $(l>0,v<0)$ and $(l<0,v>0)$ could not be
produced by material inside the Sun. Material that lies outside the Sun can
produce emission in these quadrants, but at $|l|<30\degree$, such emission
is confined to small values of $|v|$.  Velocities in the
quadrants $(l>0,v<0)$ and $(l<0,v>0)$ that cannot be reached by gas outside
the Sun are said to be  ``forbidden''.

For example, practically all the emission at $| l |<5 \degree$,
$v>100 \kms$  in Fig. \ref{fig:obs1} comes from
inside the Solar circle (but the forbidden emission extends also up to $| l |
\simeq 8\degree$). Forbidden emission is automatically generated by $\Phi_2$
\citep[e.g.][]{burtonbook,BM}.

\item{\it Velocity peaks.} At $|l|\sim2\degree$ in Fig. \ref{fig:obs1}
emission is seen at very high velocities, $|v|\simeq270\kms$. At one time
these peaks were considered evidence of a very centrally concentrated
monopole component of the Galactic potential, but since \cite{Binney++1991}
it has been widely accepted that they are generated by $\Phi_2$.

\item{\it Internal features.} The lower panel of Fig.~\ref{fig:obs1} shows a
complex pattern of ridges of enhanced emission. Fig.~\ref{fig:obs2} gives the
conventional names of the more important structures -- more comprehensive
lists of features can be found in,  for example,
\cite{Rougoor1964,Kruit1970,Cohen1975,Bania1977,Bally1988,COdata,Dame2008}.
These features would be absent if the gas distribution were axisymmetric
(lower panel of Fig.~\ref{fig:obs2}). 

\item{\it The Envelope}. The red dots in Fig. \ref{fig:obs2} trace the
envelope of the observed emission, determined as explained in \cite{SM15}.
The envelope carries the imprint of $\Phi_2$ in three ways: through the
velocity peaks, the emission at forbidden velocities, and the shoulders and
bumps that arise where an internal feature touches the envelope. 

\item{\it Bumps on the Envelope}. A mechanism by which spiral arms generate
bumps on the envelope is described in \S9.1.2 of \cite{BM}.  We identify as
particularly interesting the \emph{knee} marked with a circle in
Fig.~\ref{fig:obs2}. This knee occurs where the envelope shifts from touching
the  {arm at \it 135-km/s}  to touching the {\it connecting arm}
(green dashed and light red dashed curves in Fig.~\ref{fig:obs2}). 

\item{\it Tilt}. The central few kiloparsecs of the gas disc appear to be
tilted with respect to the Galactic plane such that the far end of the bar
lies above, and the near end below, the plane
\citep{lisztburton1980,burtonbook}. It is worth noting in the connection that
the structure of the knee changes as we shift in $b$.  In
Figs.~\ref{fig:obs1} and \ref{fig:obs2} the data are integrated over a range
in latitude $b$, but if we study slices at different latitudes
(Figs.~\ref{fig:COlatitude} and \ref{fig:HIlatitude} in Appendix
\ref{appendix:latitude}) we see that the connecting arm, which contributes to
the envelope on the low-$l$ side of the knee, appears only at $b<0$, while
the arm at $135 \kms$, which forms the envelope on the other side, appears only
at $b\geq 0$. This relative offset in $b$ suggests that the gas disc is
moving to larger $b$ as one moves outwards along the black dotted line in the
top right panel of Fig.~\ref{fig:archetypal}.  Since the connecting arm and
the arm at $135 \kms$ appear at distinct latitudes, they should represent
distinct dynamical features. Thus the knee, which arises from the transition
between these structures, cannot reflect merely a sudden change in the
circular speed. 

\item{\it Clumpiness}.  In addition to the coherent features discussed in
item (iii), the observed $(l,v)$ plots show scattered clumpiness on a variety
of scales. This clumpiness is believed to reflect clumpiness in real space,
and was modeled by \cite{Baba++2010} as caused by heating and cooling
processes.  In general, models that do not include such effects are, in the
absence of the wiggle instability, smooth on fine scales. Besides including
heating and cooling effects, proper radiative transfer modelling is probably
needed to explain the clumpiness in detail. 

\item{\it Asymmetry}. Approximately three-quarters of the molecular emission
from $|l|\la4\degree$ comes from positive longitudes
\cite[e.g.][]{burtonbook}. The cause of this asymmetry is a long-standing
puzzle. The asymmetry is too big to be
attributed solely to a perspective effect from an inclined bar
\citep{JenkinsBinney94}. The only promising explanation currently available
is that the asymmetry is generated by fluctuations in an unsteady flow
(SBM2015a). However, as mentioned in the introduction, this idea
remains embryonic and requires further investigation.

\item{\it Variation by species}. Different chemical species HI, CO, CS, etc.,
are distributed differently throughout the Galaxy because they probe
different temperature and density environments. Consequently, each produces a
different $(l,v)$ diagram. A complete model would explain the variation in $(l,v)$
plots. A related problem is the extent to which gas and dust are
correlated \citep[see for example][]{SaleMagorrian14}.

\end{enumerate}

Towards the end of the paper we will return to this list to review how
items constrain the Galactic bar.

\section{A reference model} \label{sec:archetypal}
In this Section, we describe in detail a ``reference'' model. Table
\ref{table:1} gives the values of its defining parameters. The adopted values
of the sound speed ($\cs=10\kms$) and the angle of the bar to the Sun-GC line
($\phi=20\deg$) are the same as those used by SBM15a to facilitate
comparison with earlier work. We do, however, adopt a smaller pattern speed
($\Omega_{\rm p} = 40 \kms \kpc^{-1}$) than that ($63\kms \kpc^{-1}$) used by
\cite{Binney++1991} and SBM15a for reasons that will emerge later.  The
reference model has not been chosen because its parameters give the best fit
to the observations; rather, it is just a representative model that displays
many of the characteristic of interest for this paper. In later sections, we
will see how these characteristics change as a function of the model's
parameters.

\begin{figure*}
\includegraphics[width=1.0\textwidth]{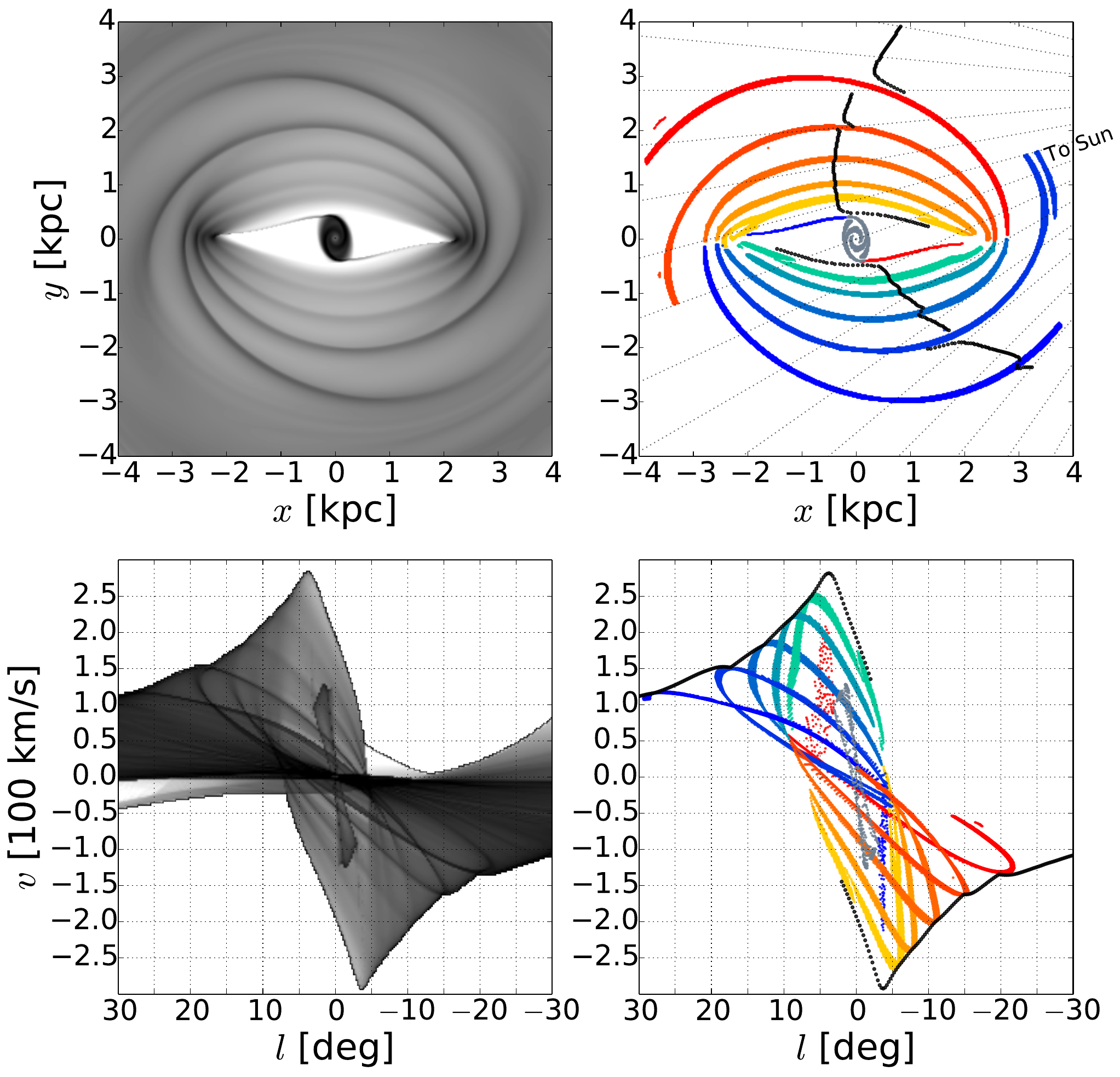}
\caption{The reference model. Top Left: the gas density in the $xy$ plane. Bottom Left: the projection of the reference model to the $(l,v)$ plane assuming an angle $\phi=20\degree$. In the two right panels the same features are shown in $xy$ and $(l,v)$ plane using the same colour coding. The black dots trace the points corresponding to the envelope in the $(l,v)$ plane.} 
\label{fig:archetypal}
\end{figure*}

The top left panel of Fig.~\ref{fig:archetypal} shows the gas density reached
after a long time. The bottom left panel of Fig.~\ref{fig:archetypal} shows
the associated $(l,v)$ plot for bar angle $\phi=20\degree$. The top right
panel marks with coloured lines prominent features in the top left panel. The
lower-right panel shows the location of these same features in the $(l,v)$
plane.

The general characteristics of the gas flow can be understood as described in
SBM15a,b and we now briefly summarise their picture. In the
outer regions, where spiral arms are present, the velocity field of the gas
is well approximated by elongated, closed ballistic $x_1$ orbits. However,
the agreement is not perfect because the streamlines execute small librations
around underlying $x_1$ orbits. These librations generate spiral arms as
kinematic density waves (SBM15b).  While flowing approximately on
$x_1$ orbits, the gas also slowly drifts inwards. At a transition point,
the gas stops following $x_1$ orbits and two straight offset shocks form.
After passing the shocks, the gas plunges towards the central $x_2$ disc,
where it settles and flows to a very good approximation on $x_2$ orbits. The
$x_2$ disc is shown as the central grey structure in the top right panel of
Fig.~\ref{fig:archetypal}, and the two straight shocks are the red and blue
features that just touch the grey structure. 

In contrast to the $(l,v)$ plot generated by circular motion (lower panel of
Fig.~\ref{fig:obs2}), the $(l,v)$ plot of the reference model (bottom left
panel of Fig. \ref{fig:archetypal}) bears a strong resemblance to the observed
$(l,v)$ plots. Because we are interested in the region inside the solar
circle, we ignore emission in the quadrants $(l>0,v<0)$, $(l<0,v>0)$ other
than emission at small $|l|$ and forbidden velocities.  The position and
strength of the velocity peaks of the reference model ($| l |\simeq
3\degree$, $v\simeq280\kms$) match well the peaks seen in the observational
plots. The region covered by forbidden velocities in the reference model
matches reasonably well that found in observations. It  is a significant
improvement over the model in SBM15a, although in the reference model the
envelope profile in the forbidden emission region is somewhat too steep. 

The envelope of the reference model displays bumps wherever the envelope is
touched by one of the lines that fan out from the centre. Each line is the
projection of a bar-driven spiral arm, as can be seen comparing the upper and
lower right-hand panels in Fig.~\ref{fig:archetypal}. The stronger a spiral
arm is, the larger the bump it creates on the envelope. 

The black dots in the upper-right panel of Fig.~\ref{fig:archetypal} show the
points in the Galactic plane that provide the emission that forms the
envelope in the $(l,v)$ plane.  Each time a spiral arm in the $(l,v)$ plane
becomes tangent to the envelope, the black dots make a discontinuous jump in
the Galactic plane.  Particularly strong bumps in the $(l,v)$ plane and large
jumps in the Galactic plane occur at $l\simeq18\degree$,$v\simeq150\kms$,
$l\simeq-15\degree$,$v\simeq-160\kms$ and
$l\simeq-20\degree$,$v\simeq-140\kms$. In Section \ref{sec:observations},
item (v), we have noted the presence of similar bumps in the observational
envelope, including  the knee circled in
Fig.~\ref{fig:obs2}. A faint bump in the reference model at
$l\simeq9\degree$,$v\simeq210\kms$ is close to the right position, but is much less
strong than the observed knee. We will see below that it can be made stronger
by varying the quadrupole. Another bump is present in the observations where
the $3\kpc$ arm touches the envelope at $| l | \simeq -18\degree$, $v =
-150\kms$. This is quite similar to the one present in the reference model at
$l\simeq-20\degree$,$v\simeq-140\kms$.

In the SBM15a model, bumps were present but were weaker than observed because
the spiral arms were barely discernible in the $(l,v)$ plane. The reference
model is a significant improvement, but its knees need to be strengthened and
the positions of its bumps tweaked.

Some of the ``arms'' in the observed $(l,v)$ plots can be identified with
features in the reference $(l,v)$ plot (Fig.~\ref{fig:archetypal}). For
example, the $3\kpc$ arm and its far-side counterparts are similar to the
outermost red arm and its counterpart on the other side, the outermost blue
arm. The connecting arm is well traced by part of the innermost green arm,
and the arm at $135\kms$ is similar to its outer neighbour. Not all arms of the
reference model have a counterpart in the observations, but it appears as if
all the principal features that contain \emph{arm} in their name have a
counterpart in the reference model.

The shocks, traced by the red and blue lines that just touch the $x_2$ disc
in the top right panel of Fig.~\ref{fig:archetypal}, are very narrow in the
Galactic plane, but when projected to the $(l,v)$ plane show quite a
spread in longitude.  This suggests that all the vertical features
identified in Figure~\ref{fig:obs2} are
different portions of the two shocks. In the Galaxy the distribution of gas
is not as smooth as in our models, and gas can be more concentrated along
regions that, when projected, produce different vertical features. According to the
picture proposed by SBM15a, the shock lanes should also form
two sides of the CO parallelogram, and the projection of the $x_2$ disc
should match the region of CS emission.  The reference model meets this
expectation in that both the size of the $x_2$ disc and the position of the
shock lanes in the $(l,v)$ plane are right. The shocks are not as prominent in
the reference $(l,v)$ plane as in the observed one, probably because the
model does not include conversion in shocks of gas to molecular form.

The molecular ring, schematically shown in Fig. \ref{fig:obs2}, is well
reproduced by our reference model, which becomes darker along a similar
diagonal band in the $(l,v)$ plane. The darkness of the molecular ring in the
$(l,v)$ plane is largely a consequence of velocity crowding: extended areas
of material outside the outermost spiral arms in the top left panel in Fig.
\ref{fig:archetypal} project to similar values of $(l,v)$ and hence produce
bright intensities in the $(l,v)$ plane. This is an improvement over the
model of SBM15a, which generates an $(l,v)$ diagram in which the dark band
runs too steeply because the gas that generates it is too centrally
concentrated, so it projects to low longitudes and high velocities.

We have seen that the reference model, qualitatively, manifests most of the
observational signatures of the bar.  The next step is to make the
correspondence between the model and observed $(l,v)$ plots quantitatively
satisfying. This proves to be a challenging task because changing the model's
physical parameters often improves the fit of one feature to the detriment of
another. In the next section we study the dependencies of features on
parameters that define the bar ($A$, $\rqu$, $\Omega_{\rm p}$).

\begin{table}
\caption{Parameters for the reference model described in detail in Sec. \ref{sec:archetypal}.}
\begin{tabular} {l c c c c c}
\toprule
\addlinespace[0mm]
	\multicolumn{1}{b{1.8cm}}{\begin{center}	model 			\vspace{0.3cm}		\end{center}} &
  	\multicolumn{1}{b{.8cm}}{\begin{center}	$A$	 			\vspace{0.3cm} 	\end{center}} &
  	\multicolumn{1}{b{.8cm}}{\begin{center}	$\rqu$			$[\kpc]$			\end{center}} &
  	\multicolumn{1}{b{1.4cm}}{\begin{center}	$\Omega_{\rm p}$		$[\kms\kpc^{-1}]$	\end{center}} &
  	\multicolumn{1}{b{.8cm}}{\begin{center}	$\cs$			$[\kms]$			\end{center}} &
  	\multicolumn{1}{b{.8cm}}{\begin{center}	$\di x$			$[\kpc]$			\end{center}} \\
\addlinespace[-2mm]
\midrule
Reference & 0.6 & 1.5 & 40 & 10 & 20 \\
\bottomrule
\end{tabular} \label{table:1}
\end{table}

\section{Effects of Varying the Quadrupole} \label{sec:quadrupole}
First we investigate changes to the potential's quadrupole, which is
characterised by three parameters: the strength $A$, the scale length $\rqu$
and the pattern speed $\Omega_{\rm p}$. The values of these parameters that we have
explored are shown in Table~\ref{table:2}.

\begin{figure*}
\includegraphics[width=1.0\textwidth]{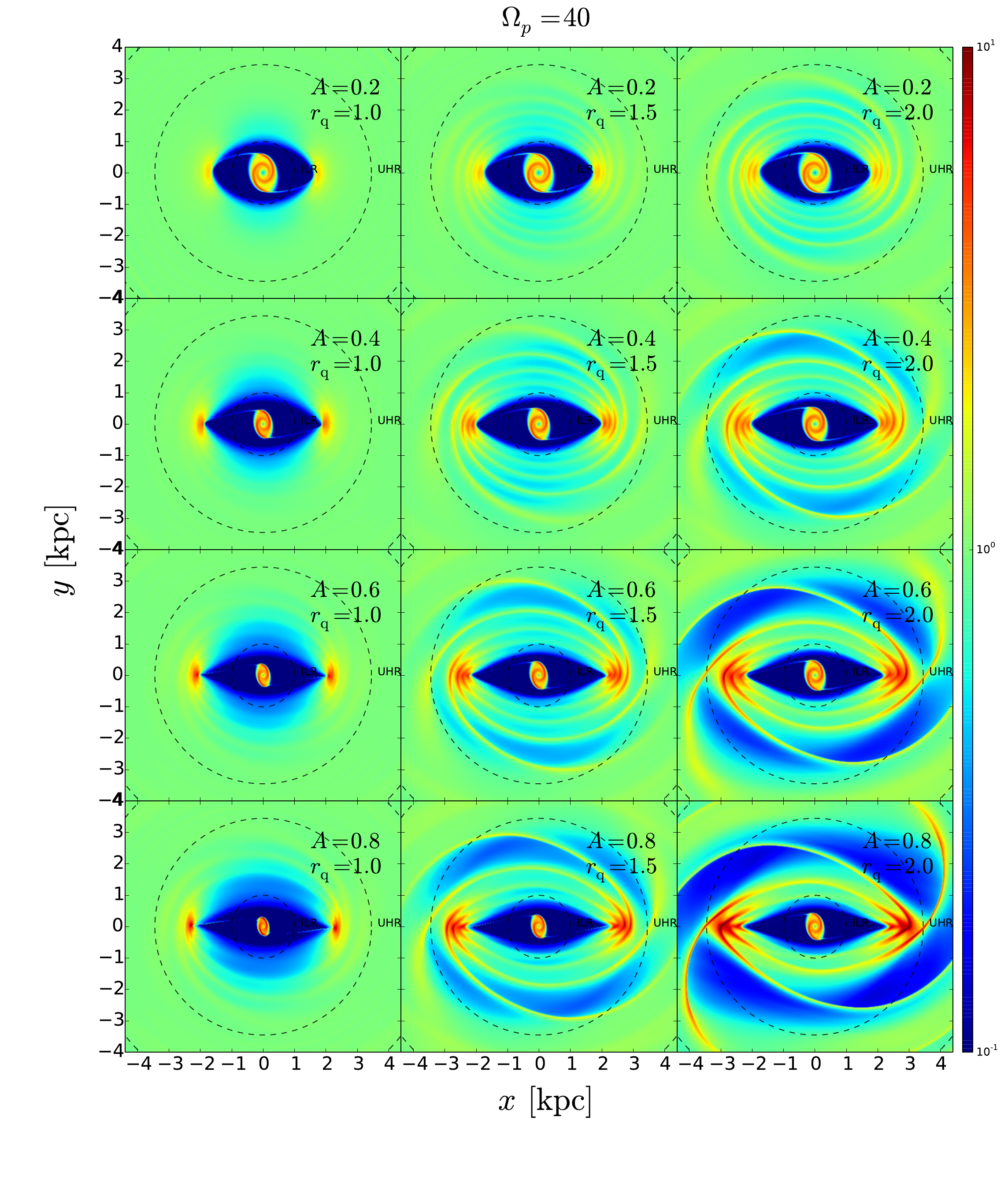}
\caption{The gas surface density in hydro simulations for different values of the quadrupole strength $A$ and quadrupole length $\rqu$. $\rqu$ is increasing left to right taking values 1.0,1.5,2.0 kpc. $A$ is decreasing from top to bottom taking values 0.2,0.4,0.6,0.8. $A$ and $\rqu$ are defined in Eq. \eqref{eq:rho2}. All snapshots are for a value of the pattern speed $\Omega_{\rm p}=40 \, \kms\kpc^{-1}$. The major axis of the bar is aligned horizontally, and gas has reached an approximately steady state in the rotating frame and circulates clockwise. All snapshots are taken at $t = 1.25\rm Gyr$. The dotted circles mark the positions of the resonances, calculated from the monopole. Since the monopole and $\Omega_{\rm p}$ are the same in all panels, the positions of the resonances are identical for all of them. The colorbar is in units of $M_\odot\pc^{-2}$} 
\label{fig:HydroOmega40}
\end{figure*}

\begin{figure*}
\includegraphics[width=1.0\textwidth]{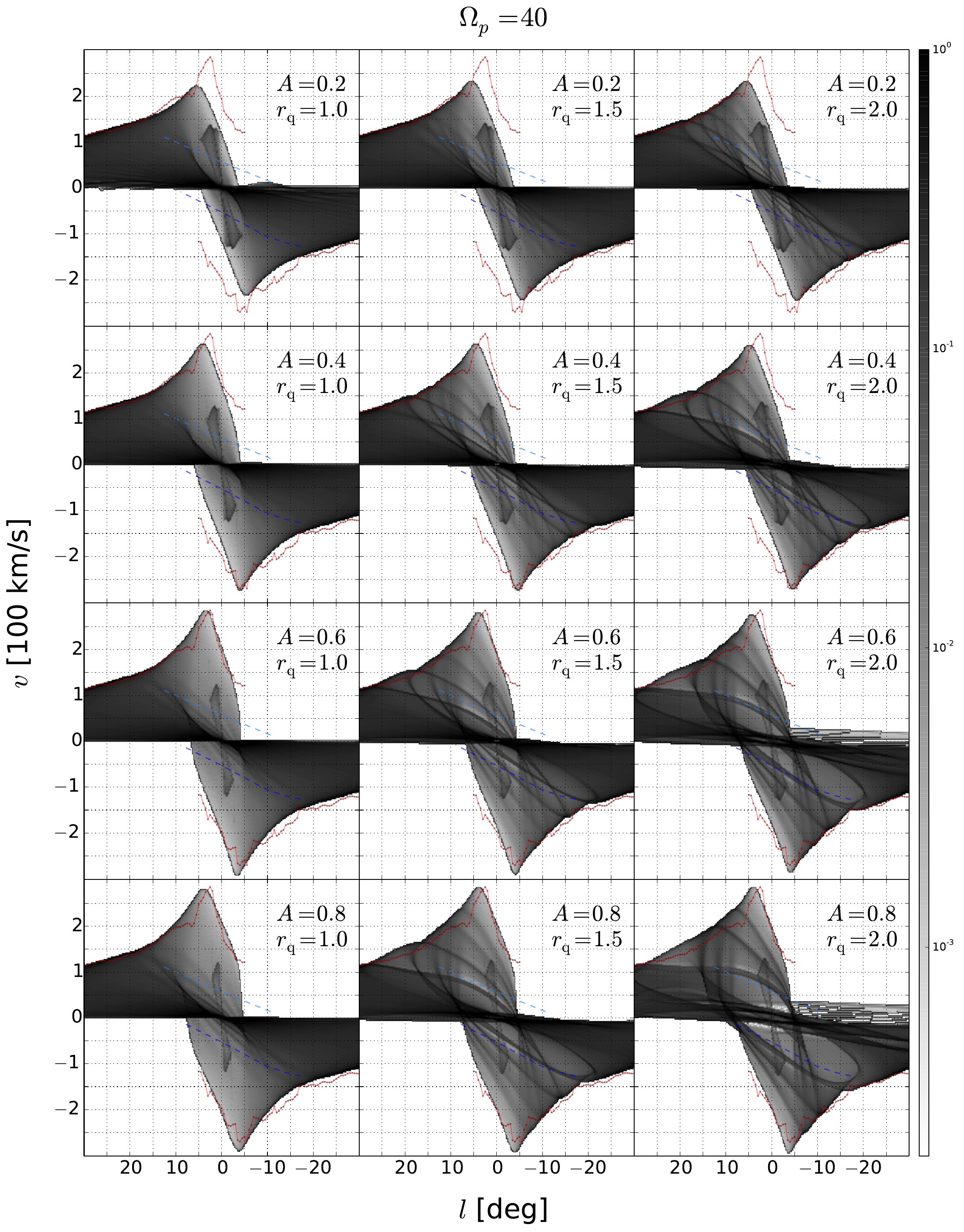}
\caption{The simulations of Fig. \ref{fig:HydroOmega40} projected into the $(l, v)$ plane. The Sun is assumed to be in a circular orbit with $v = 220 \kms$ at $R_0 = 8 \kpc$, and the bar major axis makes an angle $\phi=20\degree$ with the Sun-Galactic centre line. The red dots trace the envelope of the observations, while the blue dashed lines indicate the positions of the near and far $3\kpc$ arms. Since the projections can be scaled freely, the colorbar is in arbitrary units.} 
\label{fig:lvplot40}
\end{figure*}

\begin{figure*}
\includegraphics[width=1.0\textwidth]{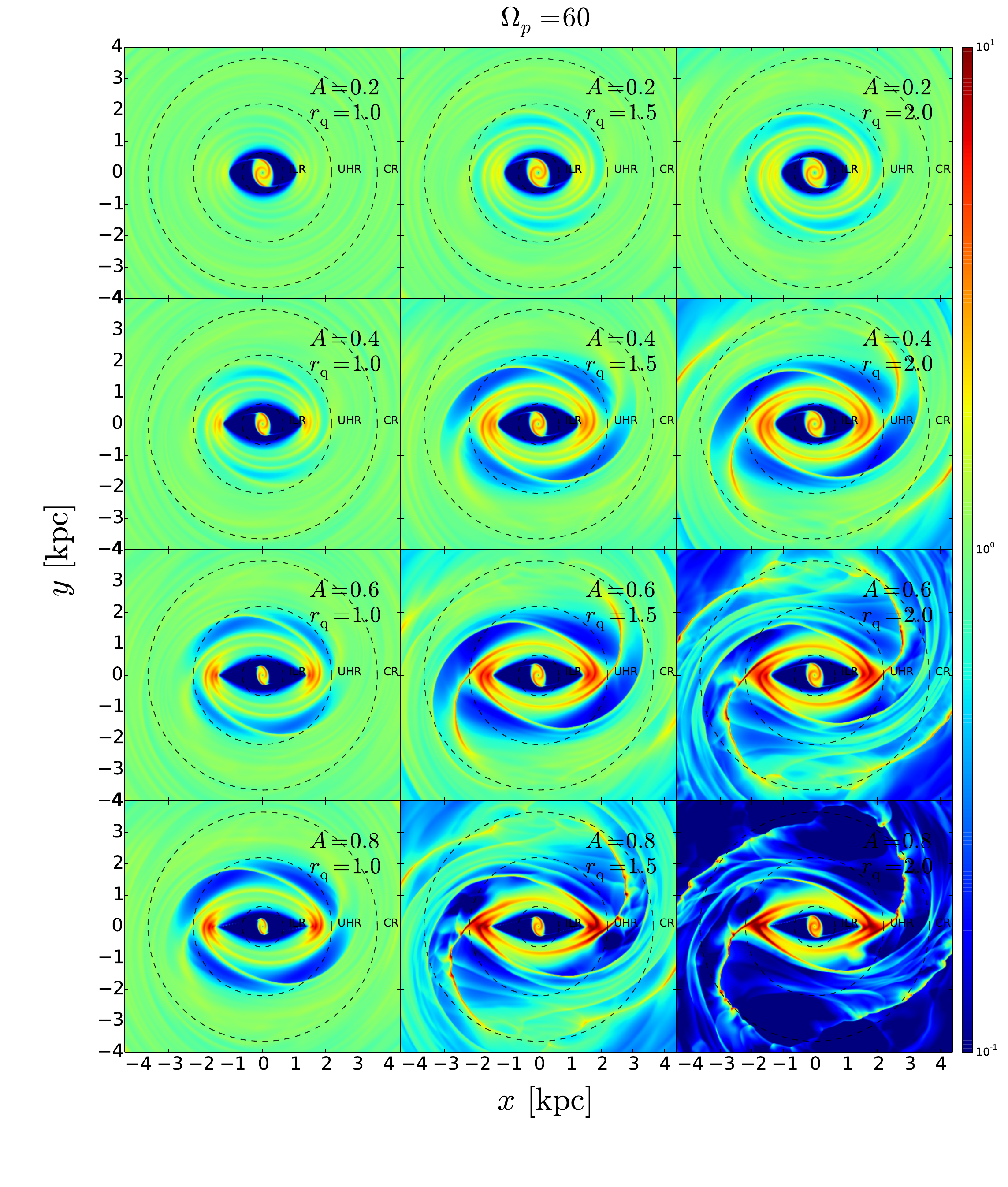}
\caption{Same as Fig. \ref{fig:HydroOmega40} but for a pattern speed $\Omega_{\rm p}=60 \, \kms\kpc^{-1}$.} 
\label{fig:HydroOmega60}
\end{figure*}

\begin{figure*}
\includegraphics[width=1.0\textwidth]{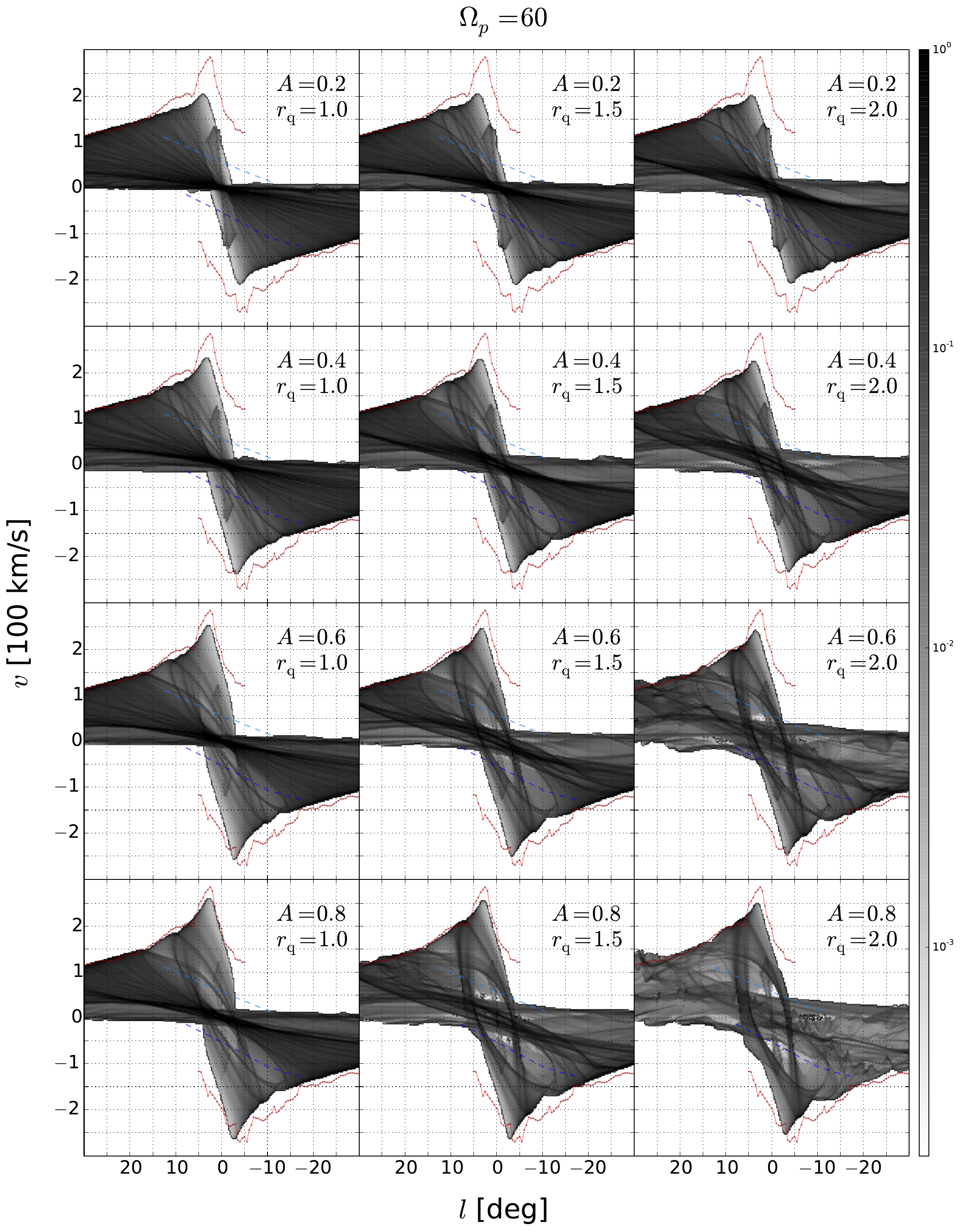}
\caption{Same as Fig. \ref{fig:lvplot40} but referring to Fig. \ref{fig:HydroOmega60}.} 
\label{fig:lvplot60}
\end{figure*}

Figs.~\ref{fig:HydroOmega40} and \ref{fig:HydroOmega60} show the gas density
for different values of quadrupole strength $A$ and length $\rqu$ and for
values of the pattern speed $\Omega_{\rm p} = 40\kms\kpc^{-1}$ and $\Omega_{\rm p} = 60\kms\kpc^{-1}$, respectively.
Figs.~\ref{fig:lvplot40} and \ref{fig:lvplot60} show the projections of the
same snapshots to the $(l,v)$ plane, assuming bar angle
$\phi=20\degree$. All snapshots are taken at $t = 1.25 \rm Gyr$, when all
simulations have reached an approximate steady state (apart from any
unsteadiness caused by the \emph{wiggle} instability). The reference model has
$\Omega_{\rm p} = 40\kms\kpc^{-1}$ and in Figs. \ref{fig:HydroOmega40} and
\ref{fig:lvplot40} lies in the panel marked with $\{A=0.6,\rqu=1.5\}$. In order
to contain the length of the main text of the paper, we have relegated the
results obtained for other values of the pattern speed to Appendix
\ref{sec:appendixscan}.

The pattern speed controls the position of the resonances: increasing the
pattern speed pushes all the resonances inwards. This is reflected in the
morphology of gas flow; many characteristics of the gas flow roughly scale
with the locations of the resonances (e.g.\ SBM15a). For example, comparing
Fig.~\ref{fig:HydroOmega40} with Fig.~\ref{fig:HydroOmega60} demonstrates
that increasing the pattern speed pushes spiral arms and the shocks inwards. 
Table~\ref{table:3} lists the position of the resonances for the explored values of the pattern speed.

From Figs.~\ref{fig:HydroOmega40} and \ref{fig:HydroOmega60} we see that the
morphology of the arms, the size of the $x_2$ disc, and the transition point
where shocks are formed also depend on both $A$ and $\rqu$. Consider the
model in the top left panel of Fig.~\ref{fig:HydroOmega40}, $\{A=0.2,\rqu=1\}$.
This model is rather flat and featureless, and this is also reflected in the
$(l,v)$ projection (top left panel of Fig. \ref{fig:lvplot40}), which does
not show anything resembling the internal features discussed in Section
\ref{sec:observations}. The model lacks high-velocity peaks, and its $x_2$ disc
is too big to coincide with observed CS emission. It is, however, likely
that higher velocity peaks and a smaller $x_2$ disc would be produced by an
increase in sound speed or in resolution (SBM2015a). 

By moving down the left column of Fig.~\ref{fig:HydroOmega40}, we increase
the bar strength $A$ at fixed bar length $\rqu=1\kpc$.  The shock region
changes causing the $x_2$ disc to shrink, but the spiral arms remain weak. In
the $(l,v)$ plane (left column of Fig.~\ref{fig:lvplot40}), the envelope of
the emission changes significantly in a way similar to what would be obtained
by an increase in sound speed or resolution, but features due to spiral arms
remain weak and the resulting diagrams are rather featureless compared to
that of the reference model (Fig.~\ref{fig:archetypal}). This happens because
spiral arms appear outside the shock region and are weak when the quadrupole
is weak there, as it is when $\Omega_{\rm p}=40\kms\kpc^{-1}$ and $\rqu=1\kpc$
(Fig.~\ref{fig:Phi2}). To strengthen the spirals we need a substantial amount of quadrupole 
in the region where they are present, hence we must either extend the
quadrupole by increasing $\rqu$, or bring the spirals in by increasing
$\Omega_{\rm p}$. The efficacy of the first strategy is illustrated by moving
horizontally in either Fig.~\ref{fig:HydroOmega40} or
Fig.~\ref{fig:lvplot40}, while the efficacy of the second strategy can be
seen by comparing equivalent panels in Figs.~\ref{fig:HydroOmega40} and
\ref{fig:HydroOmega60}.

The envelope, by contrast, depends on the flow in the vicinity of the shocks,
where the quadrupole strength peaks, and is strongly affected by change in
$A$. When the quadrupole is strong ($A=\{0.6,0.8\}$) the gas reaches high
velocities and a bigger portion of the forbidden velocity region is covered.
Hence the extent of emission at forbidden velocities strongly constrains $A$,
but barely constrains $\rqu$.

Stellar dynamics teaches us that bars cannot extend extend beyond the
corotation radius \citep[e.g.][]{Sellwood1993}, and for any given
value of $\Omega_{\rm p}$ this fact sets an upper limit on $\rqu$. The models in the 
right-hand column of Fig.~\ref{fig:HydroOmega60} violate this
constraint. These are the models in which the spiral arms are so strong that they
have become wiggle-unstable shocks.

The bumps on the envelope are strictly connected with the strength of the
spiral arms, and their positions are regulated by the parameters that
characterise the bar (see, for example, the models with $A=\{0.6,0.8\}$
in Fig. \ref{fig:lvplot40}).

One of the motivations of the present study was to improve the model in
SBM15a, which has $\Omega_{\rm p} = 63\kms\kpc^{-1}$ and a quadrupole very similar
to that obtained with $\{A=0.2,\rqu = 1.5\}$ (Fig.~\ref{fig:Phi2}).
Therefore, let us compare the SBM15a model, computed with the same spatial
resolution and sound speed, with the panels for $\{A=0.2,\rqu = 1.5\}$ in
Figs.~\ref{fig:HydroOmega60} and \ref{fig:lvplot60}. The two models are very
similar in all aspects: the spiral arms, the transition point, and the size
of the $x_2$ disc. One of the problems of SBM15a model was the lack of spiral
arms and consequently of internal features in the $(l,v)$ plane.
Fig.~\ref{fig:HydroOmega60} shows that increasing the strength of the
quadrupole produces internal features. Another problem of the SBM15a model was
that it provided insufficient coverage of forbidden velocities.
Fig.~\ref{fig:lvplot60} shows that a changing the quadrupole strength and
length do not  cure this problem. A decrease in the
pattern speed is needed. We will discuss in more detail the relation to
observation in the next section.

\begin{table}
\caption{Values of parameters explored in our study.}
\begin{tabular} {c c c}
\toprule
\addlinespace[0mm]
  	\multicolumn{1}{b{2cm}}	{\begin{center}	$A$	 							\end{center}} &
  	\multicolumn{1}{b{2cm}}	{\begin{center}	$\rqu$			$[\kpc]$			\end{center}} &
  	\multicolumn{1}{b{3cm}}	{\begin{center}	$\Omega_{\rm p}$		$[\kms\kpc^{-1}]$	\end{center}} \\
\addlinespace[-2mm]
\midrule
$\{ 0.2, 0.4, 0.6, 0.8\}$ & $\{ 1.0, 1.5, 2.0 \}$ & $\{ 20, 30, 40, 50, 60, 70\}$ \\
\bottomrule
\end{tabular} \label{table:2}
\end{table}

\begin{table}
\caption{Positions of resonances for different values of the pattern speed. ILR = Inner Lindblad Resonance, UHR = Ultra Harmonic Resonance (i.e., 4/1), CR = corotation, OLR = Outer Lindblad Resonance. These are calculated from the monopole, black curve in Fig. \ref{fig:Phi0}.}
\begin{tabular} {l c c c c}
\toprule
\addlinespace[0mm]
  	\multicolumn{1}{b{2cm}}	{\begin{center}	$\Omega_{\rm p}$		$[\kms\kpc^{-1}]$					\end{center}} &
  	\multicolumn{1}{b{1cm}}	{\begin{center}	ILR				$[\kpc]$			\end{center}} &
  	\multicolumn{1}{b{1cm}}	{\begin{center}	UHR				$[\kpc]$			\end{center}} &
  	\multicolumn{1}{b{1cm}}	{\begin{center}	CR				$[\kpc]$			\end{center}} &
  	\multicolumn{1}{b{1cm}}	{\begin{center}	OLR				$[\kpc]$			\end{center}} \\
\addlinespace[-2mm]
\midrule
20 &  2.85 & 7.35 & 10.40 & 14.25 \\
30 &  1.85 & 4.75 & 7.35 	& 10.90 \\
40 &  1.00 & 3.45 & 5.60	& 8.85 \\
50 &  0.80 & 2.70 & 4.45	& 7.40 \\
60 &  0.65 & 2.20 & 3.65 	& 6.30 \\
70 &  0.55 & 1.85 & 3.10 	& 5.45 \\
\bottomrule
\end{tabular} \label{table:3}
\end{table}

\section{Implications of the observations}\label{sec:implications}
In this section we discuss what our simulations can say about the
observations, and in particular about items listed in Section
\ref{sec:observations}.

\begin{enumerate} 

\item{\it Emission at forbidden velocities}.  The extent to which emission
extends into the forbidden zone of the $(l,v)$ diagram depends strongly on
the pattern speed, and more weakly on the strength and length of the
quadrupole.  Lower pattern speeds produce more emission in the forbidden
zone. When $\Omega_{\rm p}\gtrsim 60\kms\kpc^{-1} $, the forbidden emission is insufficiently
extended, regardless of the quadrupole's strength and length (Figs.
\ref{fig:lvplot60} and \ref{fig:lvplot70}). This statement holds even if we
allow the bar angle $\phi$ to vary. A good match with observations is
obtained at patter speeds of $30 \mhyphen 40\kms\kpc^{-1}$ (Figs
\ref{fig:lvplot30} and \ref{fig:lvplot40}). When the pattern speed is lower,
$\Omega_{\rm p}\lesssim 20\kms\kpc^{-1}$ (Fig.~\ref{fig:lvplot20}), the region is
clearly too big for bar angles $\phi\gtrsim10\degree$ that are compatible
with photometric data \citep[e.g.][]{GerhardWeggReview2014}. In light of this
finding, it is not surprising that \cite{weinersellwood1999}, who confined
their fitting to the envelope, including the forbidden-velocity region, found
$\Omega_{\rm p}\simeq42\kms\kpc^{-1}$. Thus, from the forbidden velocities alone,
our simulations would suggest a pattern speed in the range $30 \mhyphen
40\kms\kpc^{-1}$.

\item{\it Velocity peaks.} Discussion of these features is delicate because
their structure is sensitive to the spatial resolution of simulations
(SBM15a). For example, in models with weak bars (e.g. top row in
Fig. \ref{fig:lvplot60}) the peaks are not reproduced, but the results of
SBM15a show that they would be reproduced at higher resolution. The
velocity peaks depend very weakly on the bar length (Figs.~\ref{fig:lvplot40}
and \ref{fig:lvplot60}), and only marginally more strongly on the bar
strength $A$, especially once we consider the effects of resolution.
Increasing the pattern speed shifts the peaks to lower longitudes. For the
chosen angle $\phi=20\degree$, the peak at positive velocity is at the right longitude when the
pattern speed is high, $\Omega_{\rm p}\simeq60\kms\kpc^{-1}$, while it is at slightly too
high longitudes when the pattern speed is lower. The peak at negative velocity instead is at the right longitude when 
the pattern speed is low, $\Omega_{\rm p}\simeq40\kms\kpc^{-1}$, and appears at slightly too low longitudes 
when the pattern speed is higher. 

\item{\it Internal features.} Internal features provide some of the tightest
constraints on the quadrupole's length and strength. As discussed in Section
\ref{sec:quadrupole}, a significant quadrupole is needed at $\sim3\kpc$ to
produce the observed arms (e.g. left column in Fig. \ref{fig:lvplot40}). No
model with a short bar, $\rqu=1$, produces the requisite quadrupole. Shifting
the pattern speed significantly from $\Omega_{\rm p}=40\kms\kpc^{-1}$, in either
direction, shifts the region in which spirals can form away from the radial
range specified by the data (See Appendix~\ref{sec:appendixscan} for details).  The $3\kpc$ arm extends to $| l | \simeq 20
\degree$, so a substantial quadrupole is required at the arm's tangent point,
$R_0 \sin(20\degree) \simeq 3\kpc$. If the bar is this long, stellar dynamics
excludes  pattern speeds as high as $\Omega_{\rm p}=60\kms\kpc^{-1}$.  Thus,
internal features strongly suggest that the bar pattern speed is around
$\Omega_{\rm p} = 40\kms\kpc^{-1}$. This is perhaps why \cite{RFC2008}, who
used the $3\kpc$ arm as a fitting criterion, favoured a pattern speed in the
range $\Omega_{\rm p} = 30 \mhyphen 40\kms\kpc^{-1}$.  Other models with higher pattern
speeds, \citep{EnglmaierGerhard1999,Bissantz++2003} reproduce the internal
features less well. The molecular ring is also better reproduced when the pattern speed is $\Omega_{\rm p}=40\kms\kpc^{-1}$. 
These models becomes darker along the right diagonal band in the $(l,v)$ plane, similarly to the reference model in Sect. \ref{sec:archetypal}. 
When the pattern speed is higher or lower, this band becomes too much or not enough centrally concentrated respectively.

\item{\it The Envelope}. In the positive-longitude permitted quadrant, $(l>0,v>0)$, the 
envelope is well matched by simulations with high pattern speed,
$\Omega_{\rm p} = 50\mhyphen60\kms\kpc^{-1}$ (Figs.~\ref{fig:lvplot50} and~\ref{fig:lvplot60}). 
When the pattern speed is lower, (e.g., Fig.~\ref{fig:lvplot40}), the predicted envelope is generally 
higher than the observed one and the descent immediately after the peak towards larger
longitudes is too shallow; in other words, the peaks are not sharp enough. In the other permitted quadrant however, $(l<0,v<0)$,
the opposite is true and the envelope is well matched by simulations with low pattern speed, $\Omega_{\rm p} = 40\kms\kpc^{-1}$, 
while at higher pattern speed the models do not reach adequately high velocities and are too steep. 
Regarding the envelope at forbidden velocities, a better match is obtained at low pattern
speeds, as noted in item (i) above.  Thus, the envelope is sending mixed
message as regards the pattern speed. Perhaps tweaking the monopole and the bar angle $\phi$ one 
would obtain a good fit to the envelope in both the permitted
quadrants for a low pattern speed. Although the mere presence of the
quadrupole dramatically changes the envelope from its form in the
axisymmetric case, the envelope's form is surprisingly insensitive to the
quadrupole's exact length and strength.

\item{\it Bumps on the Envelope} The strength and position of the bumps in
the envelope where it is touched by a spiral arm are sensitive to the
quadrupole parameters $A$ and $\rqu$. The bump associated with the near
$3\kpc$ arm is very well reproduced at low pattern speed
(Fig.~\ref{fig:lvplot40}) when the bar is long and strong
($A\gtrsim0.6$,$\rqu\gtrsim1.5$). The knee is also close to the right
position in some of the same models -- for example, in the model
$\{A=0.8,\rqu=2.0\}$.  When the pattern speed is higher
(Fig.~\ref{fig:lvplot60}), the knee is also well reproduced in some models but
these models don't fit the $3\kpc$ arm and its bump. Changing the viewing
angle $\phi$ moves the bumps  in longitude without much change to their
structures.

\end{enumerate}
Overall the observations are best fitted when the pattern speed is
$\Omega_{\rm p}\simeq40\kms\kpc^{-1}$. Only one aspect is better explained with a
higher pattern speed of $\Omega_{\rm p}=60\kms\kpc^{-1}$: the sharpness of the positive longitude velocity peak 
and the associated portion of the envelope in the positive-velocity permitted quadrant. 
It seems likely, although it is not guaranteed, that these shortcomings can be
resolved by tweaking the monopole component of the potential, as well as the other parameters that define the bar and the gas flow. 
Moreover, the internal features strongly suggest that the bar is longer than it could
be with a high pattern speed: their extension to large longitudes requires a long quadrupole that is difficult to reconcile with the constraints from stellar dynamics. 
The presence of a long bar has also been
confirmed by photometric evidence by the recent work of
\cite{WeggGerhard2013} and \cite{WeggGerhard2015}. Hence, we favour a pattern speed
$\Omega_{\rm p}\simeq40\kms\kpc^{-1}$, in agreement with the determinations of
\cite{Fux1999}, \cite{weinersellwood1999} and \cite{RFC2008} and slightly higher than the value of $25 \mhyphen 30\kms\kpc^{-1}$ determined by \cite{Portail++2015}.

We should also mention that many other parameters influence fits to the
observations. The bar angle $\phi$ and the sound speed are particularly
relevant. Also, simulations are known to be affected by resolution effects
(SBM15a). We have neglected these parameters in this paper in part because
they have been studied in other papers
\citep[e.g.][SBM15a]{englmaiergerhard1997,PatsisAthanassoula2000}.  Had we
included them in our discussion, the parameter space to explored would have
been intractably large. As an addition, we report that we have repeated our
simulations with a higher sound speed of $\cs=20\kms$ and we have found
increasing the sound speed can reduce the number of spiral arms but does not
materially affect the principal conclusions of this paper..

Another question of interest regards the importance of self-gravity.
To test this, we have carried a small number of simulations in the
SBM15a potential taking into account the gravitational potential
generated by the gas, in addition to that of the stars that is assumed
to be externally imposed as before. We found that self-gravity of the gas is generally negligible when the 
gas density has realistic values and should be
taken into account as a refinement only after signatures of the
bar described above have been better constrained. These simulations only refers to the large-scale effects of self-gravity; 
on smaller scales not studied in this paper it is known that self-gravity is important in maintaining the structure of individual ISM clouds. 
An open issue, however, is how much the gas is affected by the spiral response of the stars
to the bar perturbation.

\section{Conclusion} \label{sec:conclusion}

The model presented in \cite{SBM2015a} had two major shortcomings. One was
the lack of internal features, that could reproduce, for example, the $3\kpc$
arm, and the other was the lack of emission at forbidden velocities. We have
run many simulations, varying systematically the quadrupole component of the
bar potential, and we have found that both these shortcomings can be cured by
adjusting the pattern speed and the quadrupole length and strength. However,
we have failed to find a set of parameters that reproduces all the important
observational features simultaneously. Good fits to individual features can
be often obtained to the detriment of other features.

Our exploration of the quadrupole parameter space suggests that the pattern
speed of the bar is around $\Omega_{\rm p} = 40\kms\kpc^{-1}$ and that the bar
exponential scale length as defined in equation \eqref{eq:rho2} must be at
least $\rqu= 1.5\kpc$, while $\rqu= 1\kpc$ is too short. The bar strength,
defined in the same equation, must be at least $A=0.4$. In our study we
haven't explored all the parameters that are important in fitting the Milky
Way: in particular, we have kept the angle between the major axis of the bar
and the Sun-Galactic centre line constant at the value $\phi=20\degree$.
Other parameters characterising the model, such as the potential's  monopole
component, and the sound speed, play an important role in determining the gas
flow. The resulting parameter space is too big to be tractable with by-eye
fitting methods. To obtain a model that can reproduce all the important
features simultaneously, automatic quantitative methods to search in
parameter space such as that described in \cite{SM15} are promising and should be explored.

\section*{Acknowledgements}

MCS acknowledges the support of the Clarendon Scholarship Fund.  
JB and JM were supported by Science and Technology Facilities Council by grants R22138/GA001 and
ST/K00106X/1. JM acknowledges support from the ``Research in Paris''
programme of Ville de Paris.  The research leading to these results has received funding from
the European Research Council under the European Union's Seventh Framework
Programme (FP7/2007-2013) / ERC grant agreement no.\ 321067.

\def\aap{A\&A}\def\aj{AJ}\def\apj{ApJ}\def\mnras{MNRAS}\def\araa{ARA\&A}\def\aapr{Astronomy \&
 Astrophysics Review}\def\apjs{ApJS}\def\apjl{ApJ}
\bibliographystyle{mn2e}
\bibliography{2d}

\appendix
\section{Longitude-velocity diagrams variation by latitude} \label{appendix:latitude}

\begin{figure*}
\includegraphics[width=1.0\textwidth]{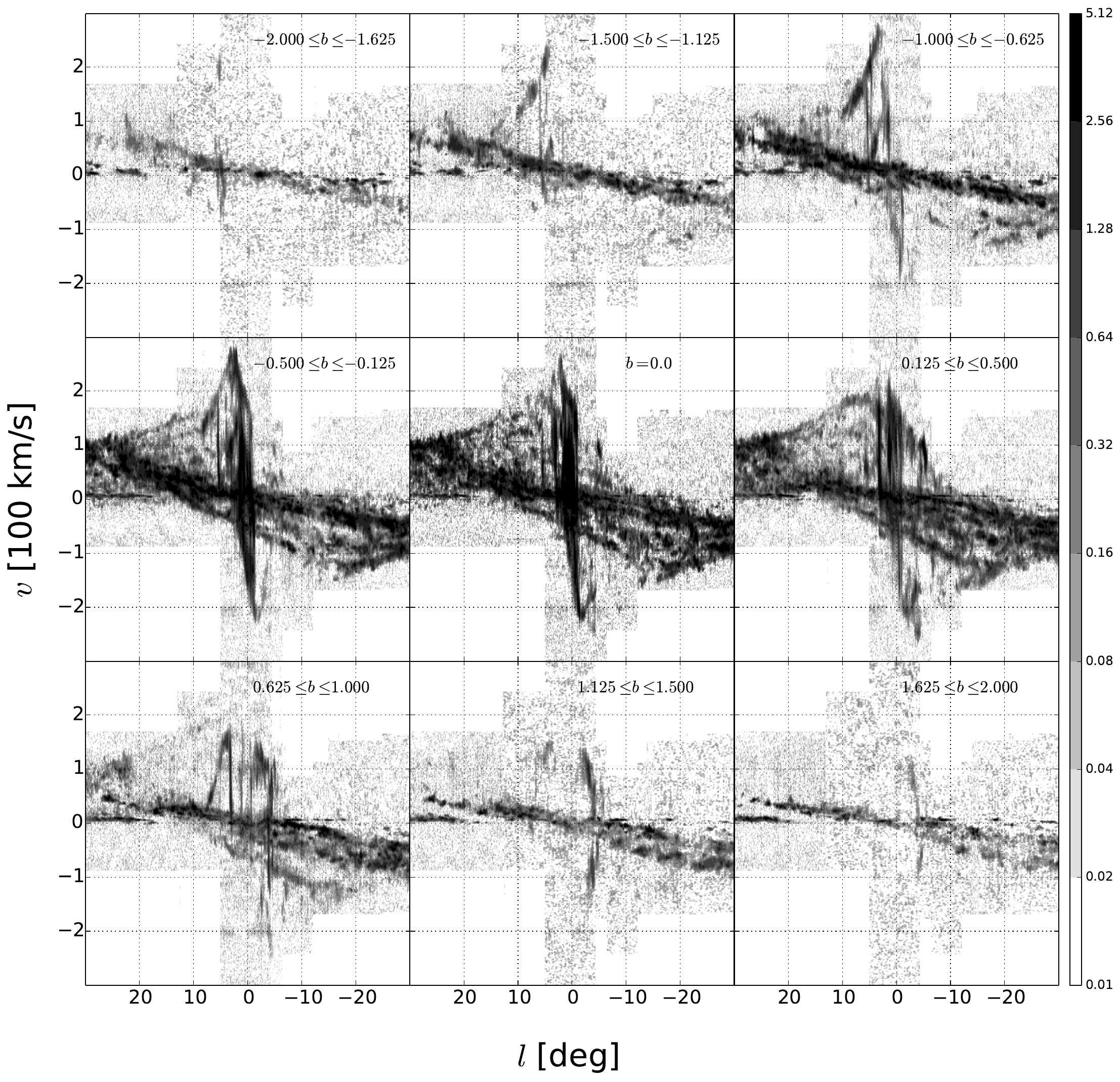}
\caption{Slices at different latitude $b$ of the same CO data shown Fig. \ref{fig:obs1}.\label{fig:COlatitude}}
\end{figure*}

\begin{figure*}
\includegraphics[width=1.0\textwidth]{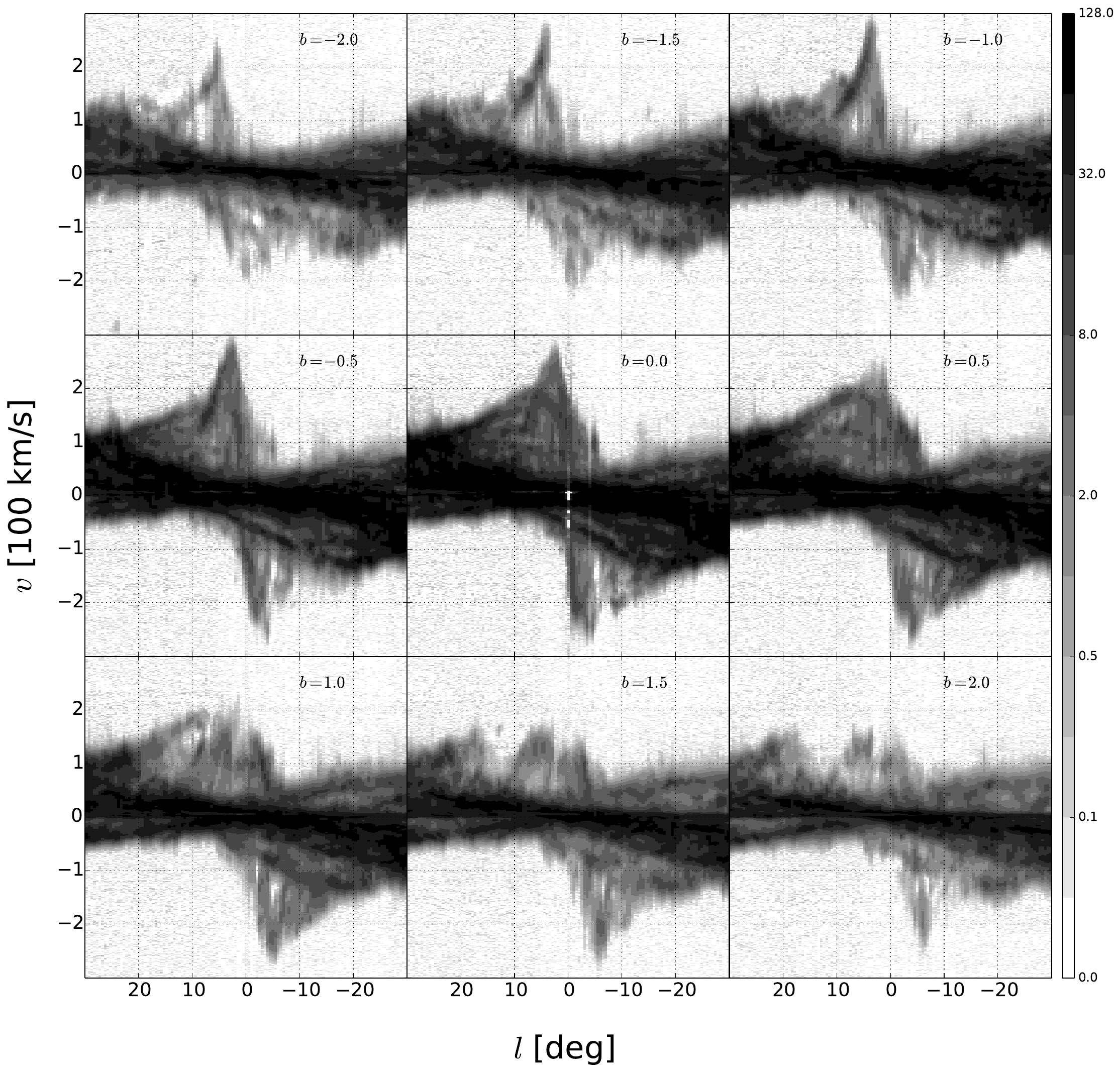}
\caption{Slices at different latitude $b$ of the same HI data shown in Fig. \ref{fig:obs1}. \label{fig:HIlatitude}}
\end{figure*}
\section{Results of the simulation for other values of the pattern speed} \label{sec:appendixscan}

In this appendix we show the results for other values of the pattern speed as a reference.

\begin{figure*}
\includegraphics[width=1.0\textwidth]{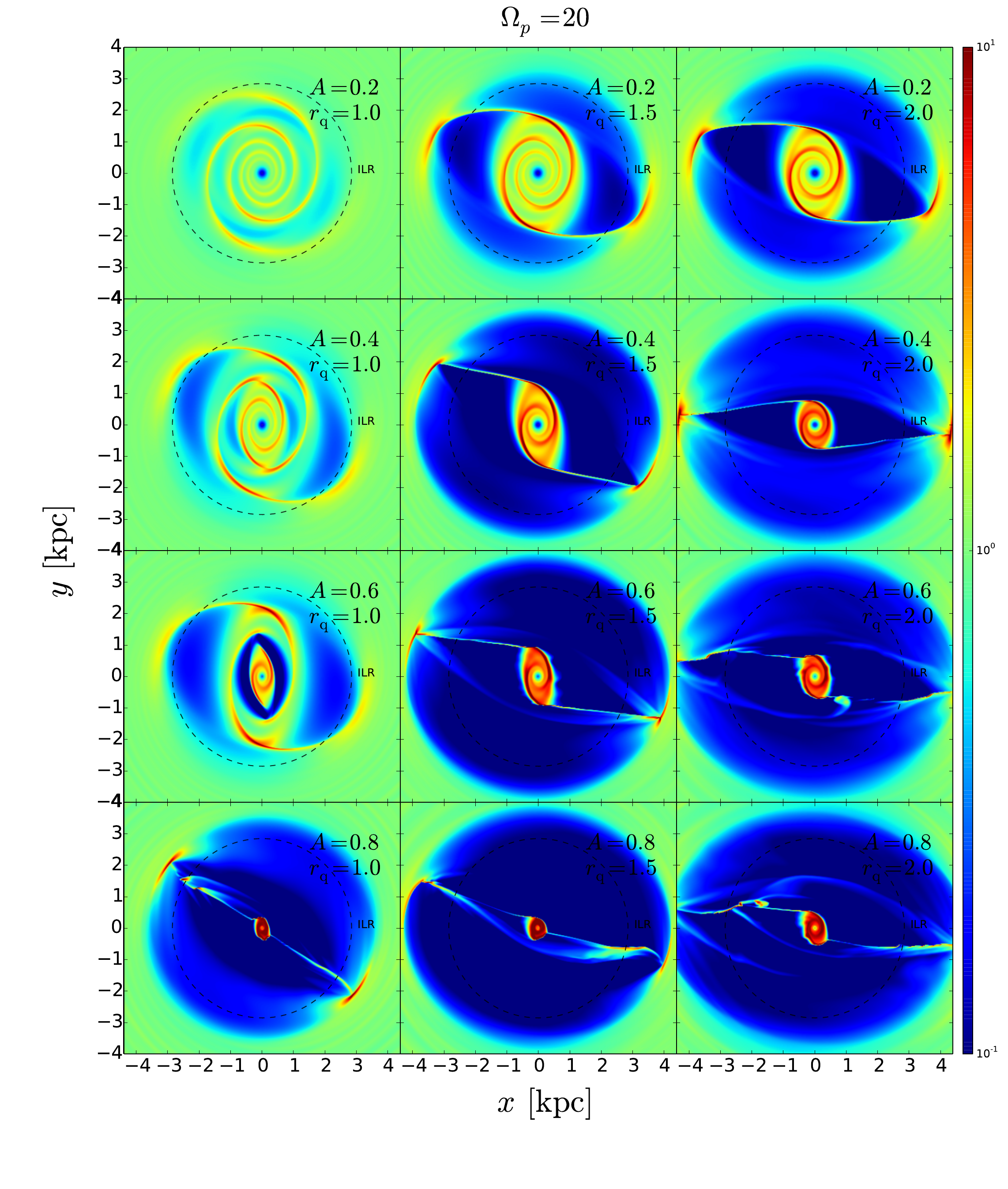}
 \caption{Same as Fig. \ref{fig:HydroOmega40} but for a pattern speed $\Omega_p=20 \, \kms\kpc^{-1}$.} 
\label{fig:HydroOmega20}
\end{figure*}

\begin{figure*}
\includegraphics[width=1.0\textwidth]{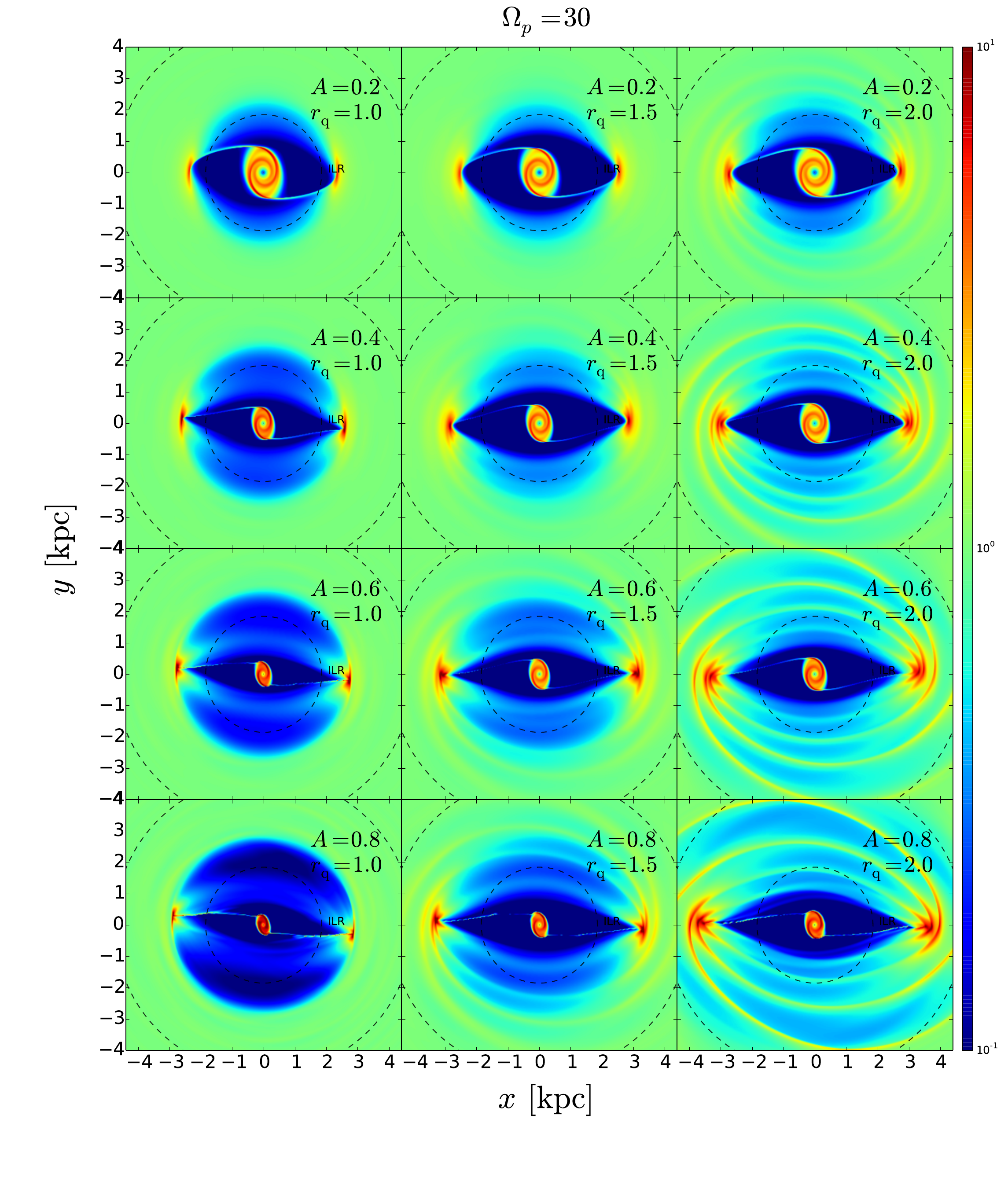}
 \caption{Same as Fig. \ref{fig:HydroOmega40} but for a pattern speed $\Omega_p=30 \, \kms\kpc^{-1}$.} 
\label{fig:HydroOmega30}
\end{figure*}

\begin{figure*}
\includegraphics[width=1.0\textwidth]{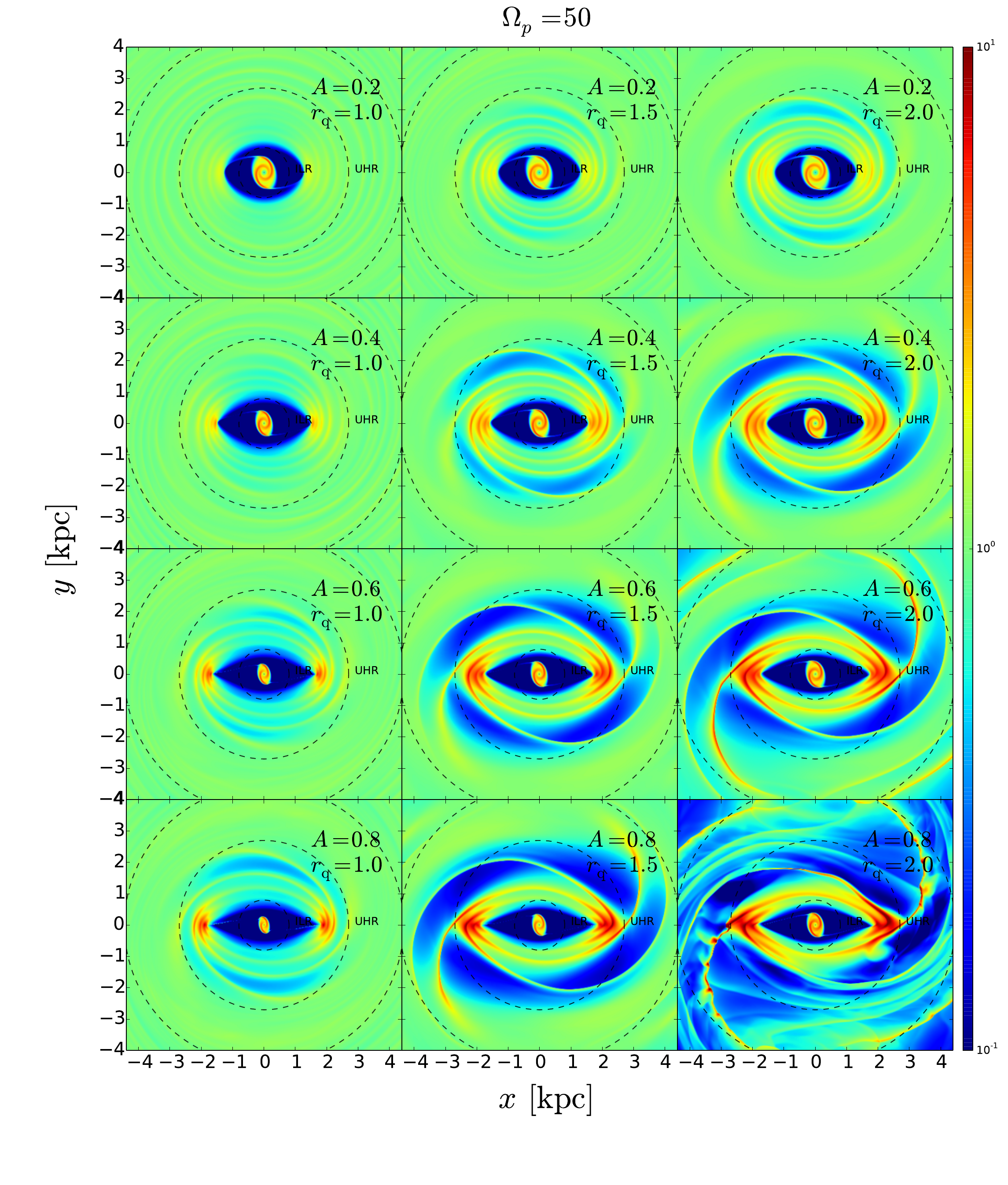}
 \caption{Same as Fig. \ref{fig:HydroOmega40} but for a pattern speed $\Omega_p=50 \, \kms\kpc^{-1}$.} 
\label{fig:HydroOmega50}
\end{figure*}

\begin{figure*}
\includegraphics[width=1.0\textwidth]{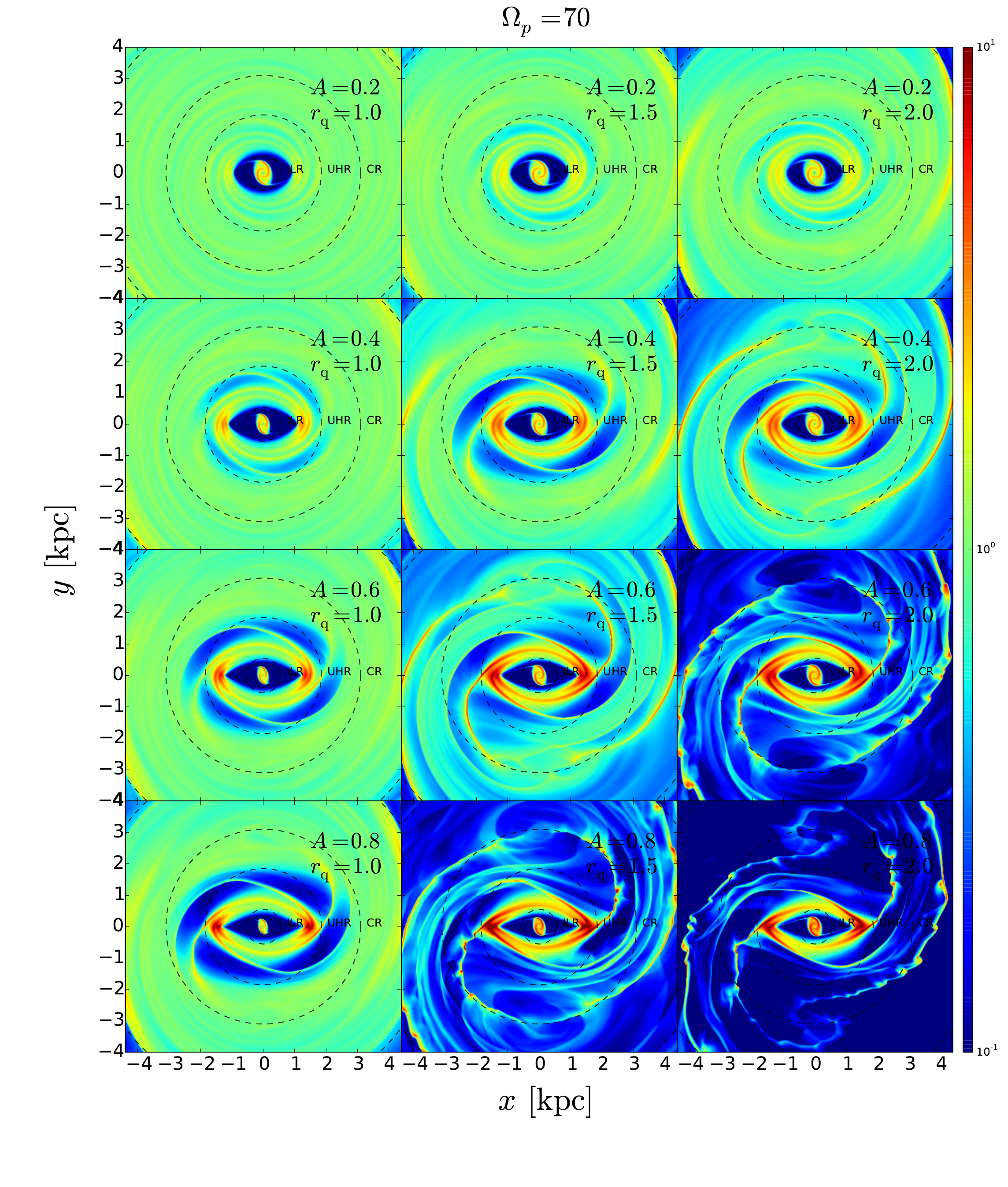}
 \caption{Same as Fig. \ref{fig:HydroOmega40} but for a pattern speed $\Omega_p=70 \, \kms\kpc^{-1}$.} 
\label{fig:HydroOmega70}
\end{figure*}

 
\begin{figure*}
\includegraphics[width=1.0\textwidth]{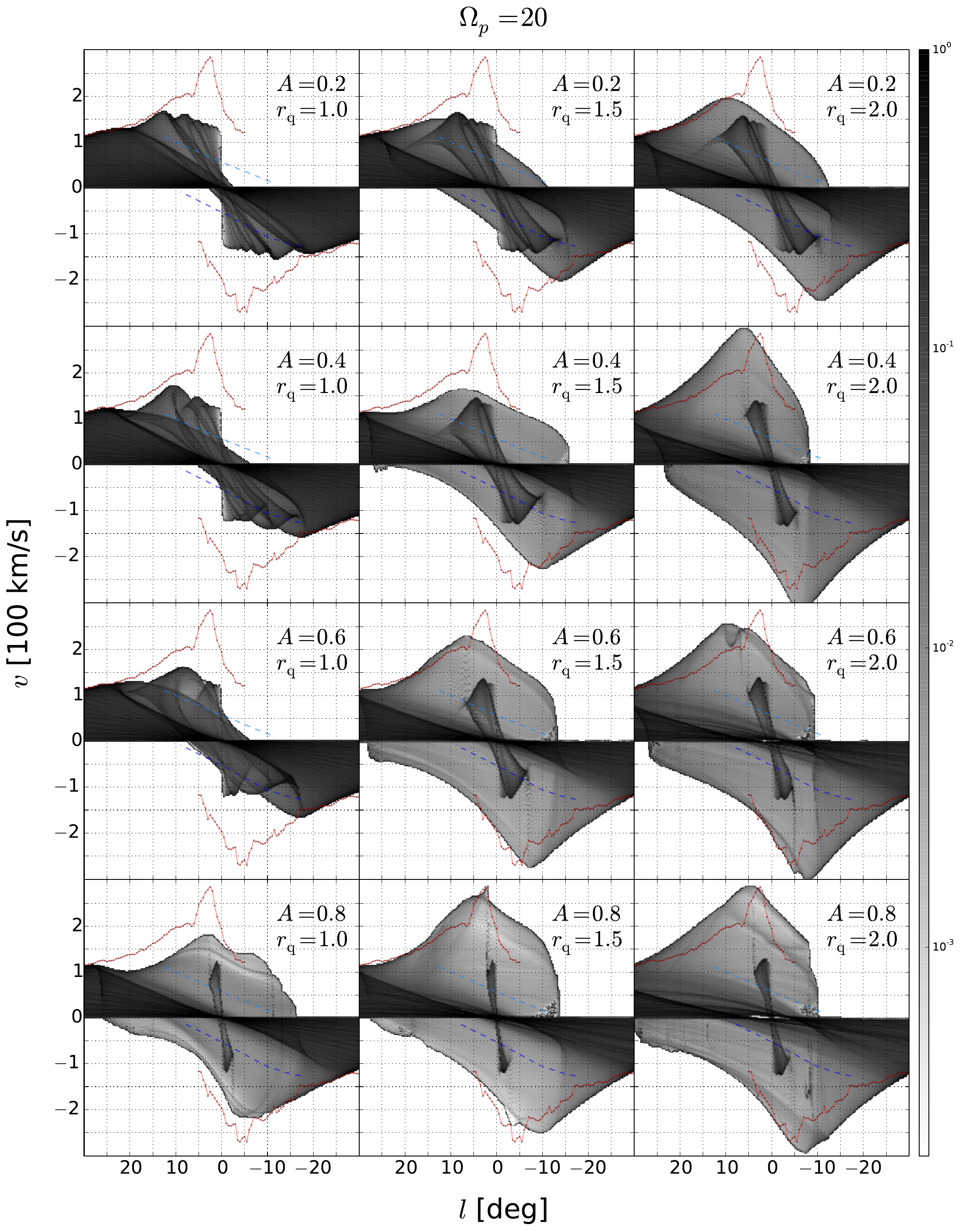}
 \caption{Same as Fig. \ref{fig:lvplot40} but referring to Fig. \ref{fig:HydroOmega20}.} 
\label{fig:lvplot20}
\end{figure*}

\begin{figure*}
\includegraphics[width=1.0\textwidth]{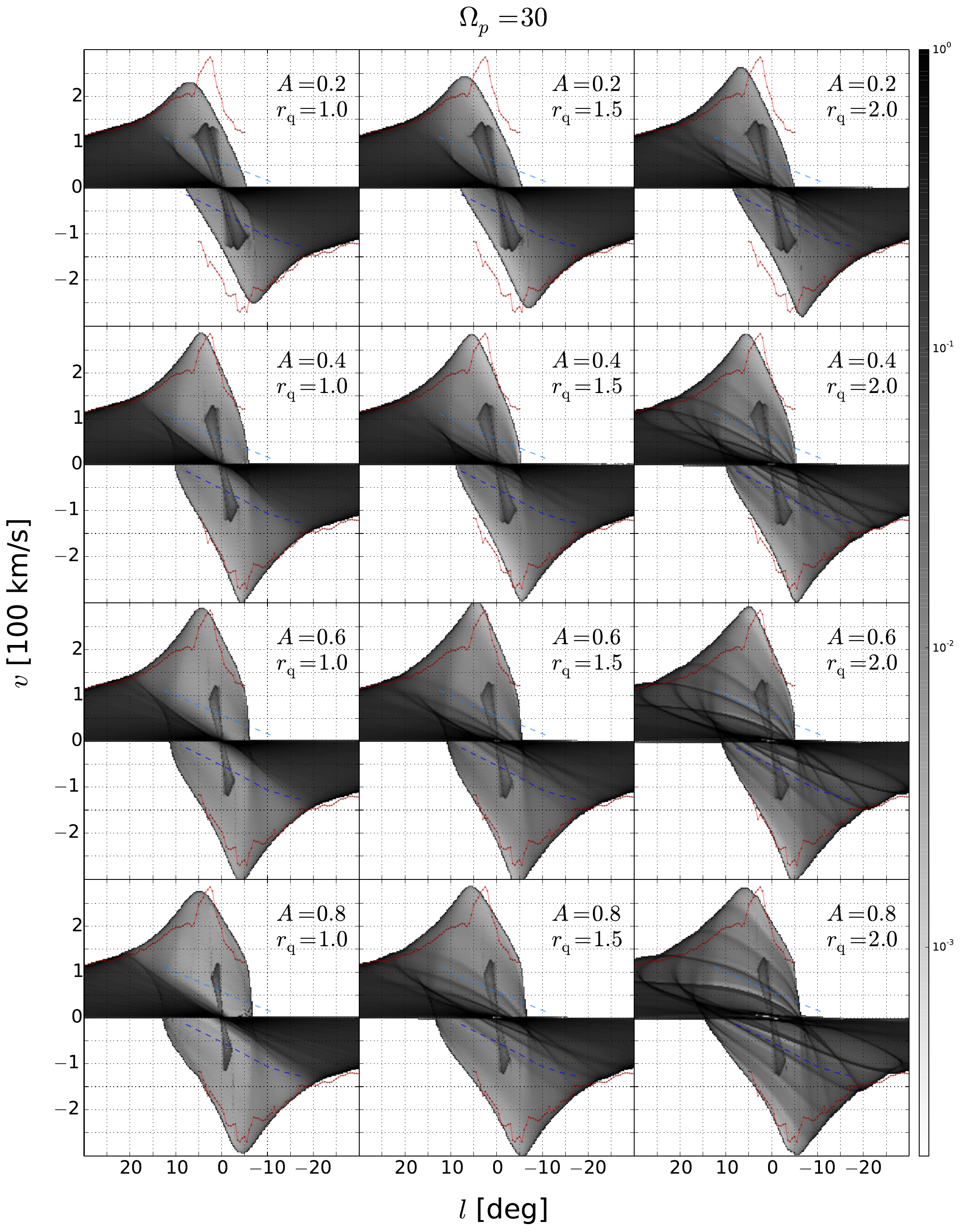}
 \caption{Same as Fig. \ref{fig:lvplot40} but referring to Fig. \ref{fig:HydroOmega30}.} 
\label{fig:lvplot30}
\end{figure*}

\begin{figure*}
\includegraphics[width=1.0\textwidth]{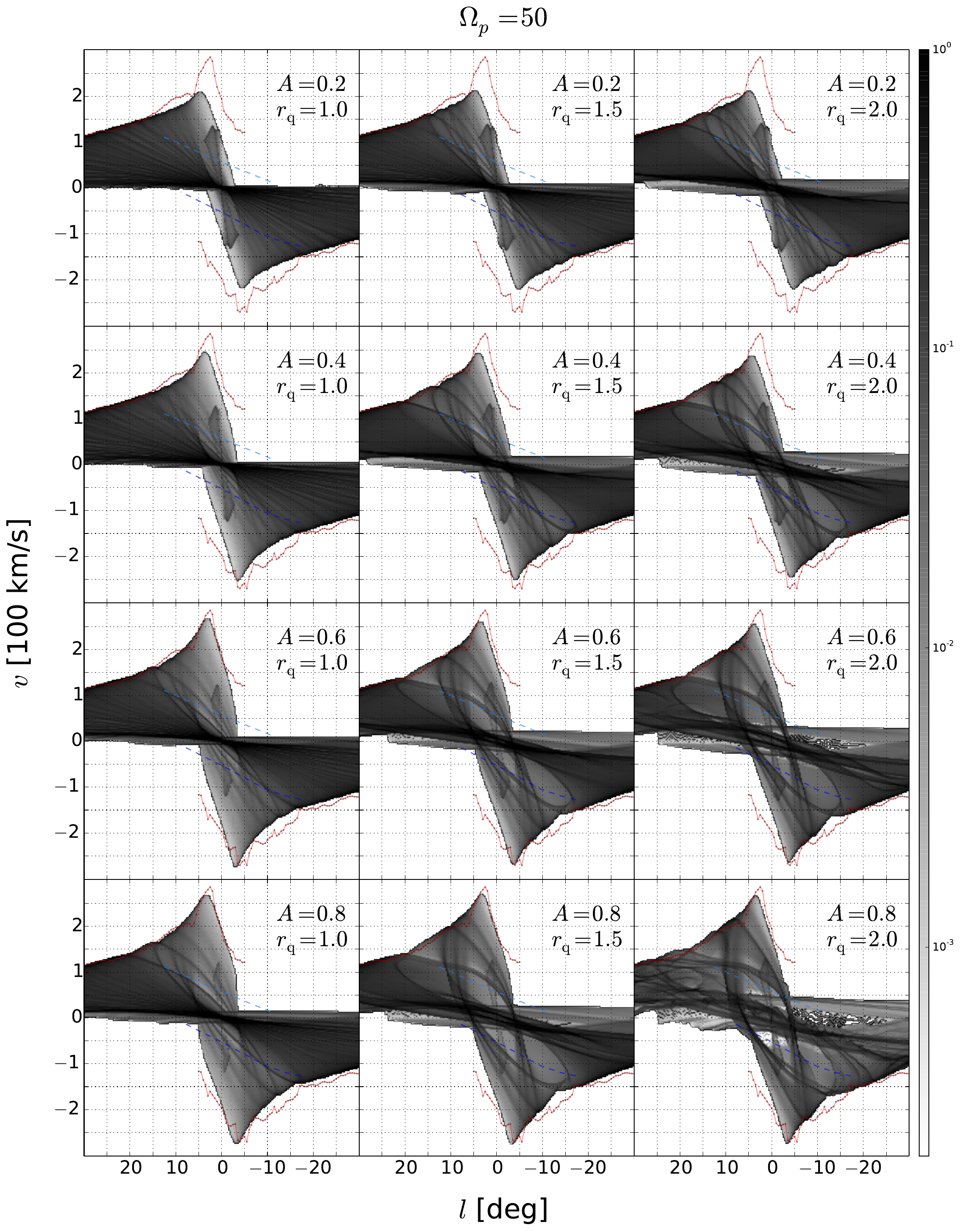}
 \caption{Same as Fig. \ref{fig:lvplot40} but referring to Fig. \ref{fig:HydroOmega50}.} 
\label{fig:lvplot50}
\end{figure*}

\begin{figure*}
\includegraphics[width=1.0\textwidth]{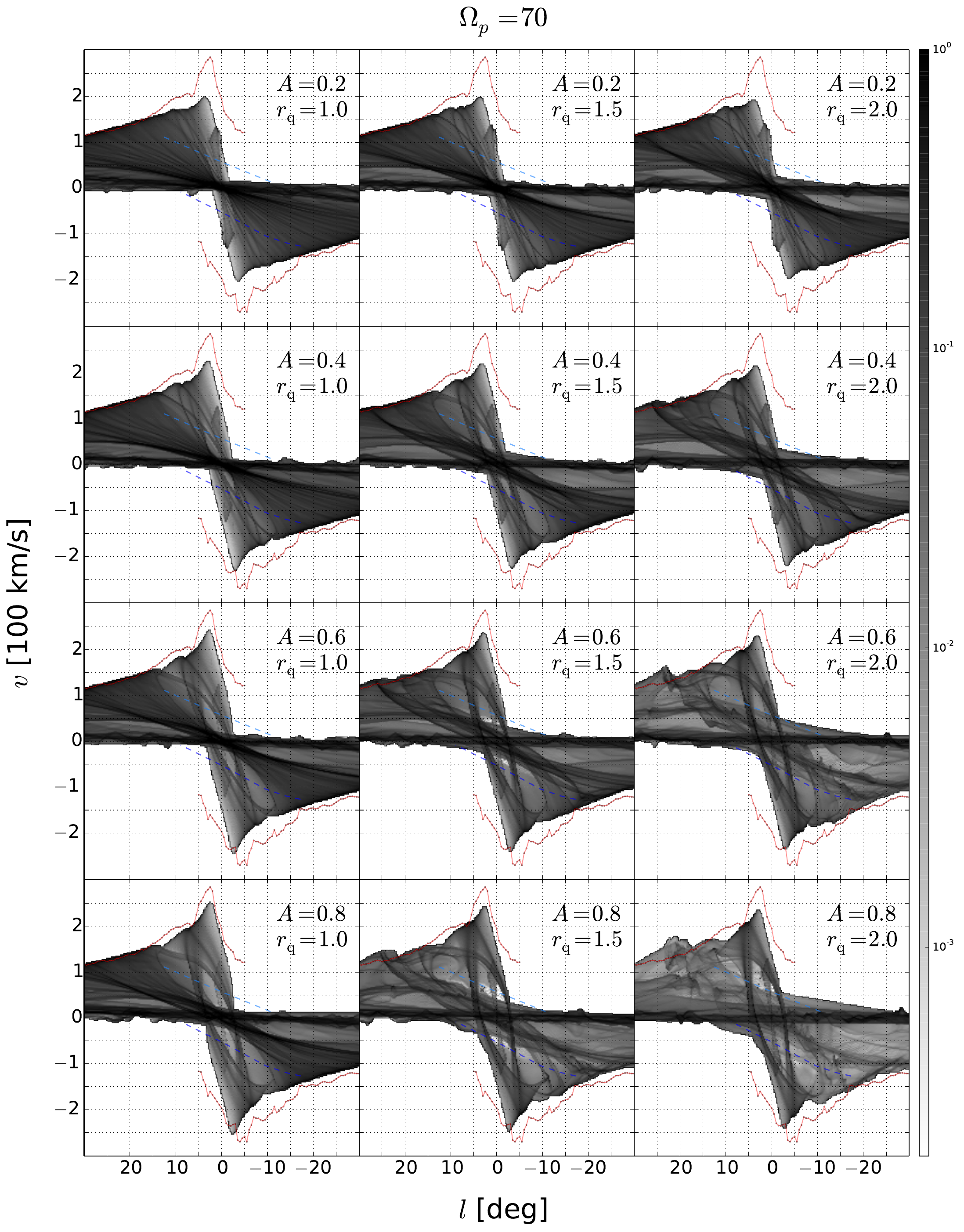}
 \caption{Same as Fig. \ref{fig:lvplot40} but referring to Fig. \ref{fig:HydroOmega50}.} 
\label{fig:lvplot70}
\end{figure*}

\end{document}